\documentclass{article}

\usepackage{microtype}
\usepackage{booktabs} 

\usepackage{hyperref}

\usepackage[font=small]{subcaption}
\usepackage[pdftex]{graphicx}
\graphicspath{{figures/}}
\DeclareGraphicsExtensions{.pdf,.jpeg,.png,.jpg}

\usepackage{amsmath}
\usepackage{amsthm}
\usepackage{amssymb}
\usepackage{xcolor}

\usepackage{url}

\newtheorem{theorem}{Theorem}[section]
\newtheorem{corollary}{Corollary}[theorem]
\newtheorem{lemma}[theorem]{Lemma}

\newtheorem{definition}[theorem]{Definition}

\newcommand{\prob}{\textsc{DenseDP}}
\newcommand{\seqalgo}{\textsc{SeqDenseDP}}
\newcommand{\palalgo}{\textsc{ParDenseDP}}
\newcommand{\phasealgo}{\textsc{PhaseDenseDP}}
\newcommand{\mralgo}{\textsc{MRDenseDP}}
\newcommand{\Graphs}{\mathcal{G}}
\newcommand{\D}{\mathcal{D}}
\newcommand{\R}{\mathcal{R}}
\newcommand{\T}{\mathcal{T}}

\newcommand{\degree}[2]{\text{deg}_{S_t}^{#1}(#2)}
\newcommand{\jAdjE}{{j \in \text{adj}(e)}}
\newcommand{\jnotAdjE}{{j \notin \text{adj}(e)}}

\usepackage[accepted]{icml2021}

\icmltitlerunning{Differentially Private Densest Subgraph Detection}

\begin{document}

\twocolumn[
\icmltitle{Differentially Private Densest Subgraph Detection}



\icmlsetsymbol{equal}{*}

\begin{icmlauthorlist}
\icmlauthor{Dung Nguyen}{uvacs,bio}
\icmlauthor{Anil Vullikanti}{uvacs,bio}
\end{icmlauthorlist}

\icmlaffiliation{uvacs}{Department of Computer Science, University of Virginia, Virginia, USA}
\icmlaffiliation{bio}{Biocomplexity Institute and Inintiative, University of Virginia, Virginia, USA}

\icmlcorrespondingauthor{Dung Nguyen}{dungn@virginia.edu}

\icmlkeywords{Machine Learning, ICML}

\vskip 0.3in
]



\printAffiliationsAndNotice{}  
\begin{abstract}
Densest subgraph detection is a fundamental graph mining problem,  with a large number of applications. There has been a lot of work on efficient algorithms for finding the densest subgraph in massive networks. However, in many domains, the network is private, and returning a densest subgraph can reveal information about the network. Differential privacy is a powerful framework to handle such settings.
We study the densest subgraph problem in the edge privacy model, in which the edges of the graph are private. We present the first sequential and parallel differentially private algorithms for this problem. We show that our algorithms have an additive approximation guarantee.
We evaluate our algorithms on a large number of real-world networks, and observe a good privacy-accuracy tradeoff when the network has high density. 
\end{abstract}
\section{Introduction}
\label{sec:intro}

Data privacy is a fundamental challenge in many real world applications, e.g., healthcare, social networks, and finance, where there is a risk of revealing private information through adversarial queries. Differential privacy (defined in Section \ref{sec:prelim}), developed through the work of a number of researchers, e.g., \cite{dwork2011differential, 10.1145/1065167.1065184, dwork:fttcs14, 10.1145/1250790.1250803}, has proven to be a very powerful approach to support diverse kinds of computations with rigorous privacy guarantees (see \cite{zhu:tkde17, Vadhan2017} for extensive surveys on this topic). This framework allows database owners to support queries with a very controlled and rigorous loss of privacy. Differentially private algorithms have now been designed for a number of problems, including supervised and unsupervised machine learning, social network analysis and deep learning \cite{Kasiviswanathan:2013:AGN:2450206.2450232, blocki:itcs13, 10.1145/2976749.2978318, 10.1145/3219819.3226070}. However, network problems have been proven to be much harder, and private algorithms for many problems (e.g., in graph mining) remain poorly understood. For some of the problems which have been studied, e.g.,~\cite{nguyen:comm16}, good accuracy bounds are not known. Further, most private algorithms do not scale very well, especially for network problems, e.g.,~\cite{Kasiviswanathan:2013:AGN:2450206.2450232, blocki:itcs13}, and there has been limited work on parallel algorithms.

Here, we study the problem of finding the densest subgraph in a graph $G(V,E)$, which is a very basic subroutine in graph mining, and has been used in diverse domains, including bioinformatics, network science, fraud detection, and social network analysis, e.g.,~\cite{cadena:procieee18,hooi:kdd16,khuller:icalp09,cadena:icdm16,tsourakakis2013denser,rozenshtein2014event}. 
There are many notions of density~\cite{cadena:procieee18}; here, we focus on the notion of average density $\rho(S)$ of  $S\subset V$, defined as $\rho(S) = (\mbox{\#edges with both end points in $S$})/|S|$. The goal is to find a densest subgraph $S^* = \text{argmax}_{S\subseteq V}\rho(S)$; we use $\rho(G)=\rho(S^*)$ to indicate the density of the the densest subgraph in $G$. This notion of density is one of the most common studied ones, e.g., for anomaly detection in networked data~\cite{cadena:procieee18,hooi:kdd16}, primarily because it can be computed very efficiently~\cite{charikar:dense,asahiro2002complexity,khuller:icalp09,goldberg1984finding}. In particular, the densest subgraph can be computed optimally using linear programming~\cite{goldberg1984finding, charikar:dense}, and a simple iterative greedy algorithm gives a $\frac{1}{2}$-approximation \cite{charikar:dense,asahiro2002complexity}. 
Highly scalable parallel algorithms have also been developed, e.g.,~\cite{bahmani2012densest, pmlr-v97-ghaffari19a}. However, private algorithms are not known for the dense subgraph problem, which is the focus of this paper. 

There are two standard approaches for privacy in networks---edge privacy (in which the nodes are public and the edges are private) and node privacy (in which the nodes are also private). As we observe later, the maximum density value has low sensitivity can be computed easily using the Laplace mechanism. However, in graph mining settings, analysts are interested in finding the actual densest subgraph (i.e., the nodes in the subgraph), which is the focus of our paper, and is a much harder problem. It only makes sense in the edge privacy model. 
Our contributions are summarized below.
\begin{itemize}
\item 
We first present a sequential $(\epsilon, \delta)$-differentially private algorithm (\seqalgo) for the densest subgraph problem in the edge privacy model. We adapt Charikar's algorithm~\cite{charikar:dense}, using the exponential mechanism~\cite{mcsherry:focs07, gupta2010} iteratively, and prove that our mechanism achieves the same multiplicative approximation factor as \cite{charikar:dense}, with an additive logarithmic factor, i.e., the solution $S$ computed by \seqalgo{} satisfies $\rho(S)\geq \rho(S^*)/2 - O(\log{n})$, with high probability.
\item
We alter the sampling process in \seqalgo{}, and obtain a parallel algorithm (\palalgo), which involves sampling multiple nodes in parallel, in each iteration. This works well in practice, in some regimes, but can have exponential time in the worst case. Our final algorithm (\phasealgo) groups multiple iterations into a single sampling phase (basically guessing when a node would be sampled). We prove that \phasealgo{} takes $O(\log{n})$ phases, with high probability. Both \palalgo{} and \phasealgo{} have similar accuracy as \seqalgo{}, though the constant factors get worse.
We also design a distributed version of \phasealgo{} in the Map-Reduce model~\cite{dean2008mapreduce, karloff:soda10, beame:pods13}. We show it has similar accuracy and runtime guarantees as \phasealgo{}.
\item 
We evaluate our algorithms on a large number of diverse networks, in order to understand the privacy-accuracy tradeoffs. We find that in networks with high densities, our private algorithms have good accuracy, especially for larger values of $\epsilon$. 
Our algorithms also have reasonable recall, i.e., the private solution contains a large fraction of the optimal subgraph. \seqalgo{} has better accuracy than \palalgo{}, which, in turn, is better than \phasealgo{}. In some regimes of $\epsilon$, \palalgo{} has significantly better accuracy than \phasealgo{}, and is comparable to \seqalgo{}.

\item 
We show an additive lower bound of $\Omega(\sqrt{\log{n}})$ for any differentially private algorithm for \prob{}.
\end{itemize}

Due to the limited space, we have omitted many proofs and other discussion; these are presented in the Appendix. The full version including the Appendix is available at~\cite{DBLP:journals/corr/abs-2105-13287}. We note that our analysis is not very tight, in order to simplify the discussion; we expect the constant factors can be tightened with a more careful analysis.


We use the exponential mechanism~\cite{mcsherry:focs07} in an iterative manner, which involves choosing elements from a set with probability that is proportional to a score associated with the element. This approach has been used in private versions of other combinatorial optimization problems, e.g., vertex cover, global minimum cut and set cover, e.g.,~\cite{gupta2010}. However, this is inherently sequential, and we develop a parallel version, in which multiple elements can be chosen simultaneously. Consequentially, the analysis of the performance becomes much more challenging, and is one of our contributions.


\section{Preliminaries}
\label{sec:prelim}

\subsection{Problem statement}

\noindent
\textbf{Densest subgraph.}
Let $G=(V, E)$ denote an undirected graph. For a subset of nodes $S\subset V$, let $E[S]$ denote the set of edges with both end points in $S$. The density of a set $S$ in $G$ is defined as $\rho(S)=\frac{|E[S]|}{|S|}$~\cite{charikar:dense,khuller:icalp09}. The objective is to find a subset $S^*(G)=\mbox{argmax}_{S\subset V} \rho(S)$, which achieves the maximum density; we refer to it as $S^*$ when the graph is clear from the context.
Let $\rho(G) = \rho(S^*)$.
We say that $\rho(S)$ is an $\alpha$-approximate solution if $\rho(S)\geq\alpha\rho(S^*)$.
An optimal densest subgraph (i.e., $\alpha=1$) can be computed optimally using a linear programming based algorithm, and an iterative algorithm gives a $\alpha=1/2$-approximation~\cite{charikar:dense,khuller:icalp09}. However, this turns out to be difficult to achieve under privacy, and we consider a slightly relaxed version: we say $S$ is an $(\alpha, \beta)$-approximation if $\rho(S)\geq \alpha\rho(S^*(G)) - \beta$.
Let $N^G(v)$ denote the set of neighbors of $v$ in $G$, and let $\deg^G(v)=|N^G(v)|$ denote the degree of $v$. 
For a subset $S\subset V$, let $\deg^G_S(v)=|N(v)\cap S|$ denote the number of neighbors of $v$ in $S$ in graph $G$; when $G$ is clear from the context, we denote it by $\deg_S(v)$.




\noindent 
\textbf{Differential privacy on graphs.} Let $\Graphs$ denote a set of graphs on a fixed set $V$ of nodes. For a graph $G\in\Graphs$, we use $V(G)$ and $E(G)$ to denote the set of nodes and edges of $G$, respectively.
In this paper, we will focus on the notion of \emph{Edge privacy}~\cite{blocki:itcs13}, where all graphs $G\in\mathcal{G}$ have a fixed set of nodes $V(G)=V$, and two graphs $G, G'\in \Graphs$ are considered neighbors, i.e., $G\sim G'$, if they differ in exactly one edge, i.e., $|E(G)-E(G')|=1$. We note that another notion that has been considered is
\emph{Node privacy}~\cite{Kasiviswanathan:2013:AGN:2450206.2450232}: two graphs $G, G'\in\Graphs$ are considered neighbors, i.e., $G\sim G'$, if they differ in exactly one node, i.e., $|V(G)-V(G')|=1$. Note that they might differ in many edges.

\begin{definition}
\label{def:dp}
A (randomized) algorithm $M:\Graphs\to R$ is  $(\epsilon, \delta)$-differentially private if for all subsets $S\subset R$ of its output space, and for all $G, G'\in \Graphs$, with $G\sim G'$, we have $Pr[M(G) \in S] \leq e^{\epsilon} Pr[M(G') \in S] + \delta$~\cite{dwork:fttcs14, dwork2011differential, Vadhan2017, blocki:itcs13}.
\end{definition}




\noindent
\textbf{Problem statement: densest subgraph detection with edge differential privacy (\prob{})} Given a family of graphs $\mathcal{G}$ on a set $V$ of vertices, and parameters $\epsilon, \delta$, construct an $(\epsilon, \delta)$-differentially private mechanism $M:\mathcal{G}\rightarrow 2^V$, such that: (1) for any graphs $G\sim G'$, and $\mathcal{S}\subset 2^V$, $\Pr[M(G)\in\mathcal{S}] \leq e^{\epsilon}\Pr[M(G')\in\mathcal{S}] + \delta$, and (2) $\rho(M(G))$ is maximized.

We evaluate the accuracy of the mechanism $M$ (in the second condition above) in terms of the $(\alpha,\beta)$-approximation to $\rho(S^*(G))$ for any graph $G\in\mathcal{G}$; therefore, we would like $\alpha$ to be as large as possible, and $\beta$ to be as small as possible. 

\noindent
\textbf{\emph{Computing density value vs finding a subgraph.}} The \prob{} problem involves finding a subset $M(G)\subset V$, where $V$ is not private, but only the edges are private. It can be verified easily that the `sensitivity''  (Definition~\ref{def:sensitivity}) $\Delta_{\rho} = \max_{G\sim G'} |\rho(S^*(G)) - \rho(S^*(G'))| \leq 1$, as the addition or removal of an edge can be shown to alter the density by at most $1$. As a result, private computation of $\rho(S^*(G))$ can be done easily with
the standard Laplacian mechanism $Lap(1)$~\cite{dwork:fttcs14}. However, returning a subgraph with high density, and with privacy is a harder problem.

\subsection{Additional background on differential privacy}
\label{sec:dp-basic}
\noindent
Let $A_1:\D_1\to R_1$ and $A_2:\D_2\to R_2$ be $(\epsilon_1, \delta_1)$ and $(\epsilon_2, \delta_2)$-differentially private algorithms, respectively. Let $f:R_1\to R'_1$ be an arbitrary randomized mapping.
We will make extensive use of the following basic results shown in \cite{dwork:fttcs14}, that differential privacy is preserved under:\\ 
(1) post-processing (Proposition 2.1 of~\cite{dwork:fttcs14}), i.e., $f\circ A_1: \D\to R'_1$ is also $(\epsilon, \delta)$-differentially private, and\\ 
(2) composition (Theorem 3.16 of~\cite{dwork:fttcs14}), i.e., 
$A:(D_1,D_2)\to(R_1,R_2)$ defined as $A(x) = (A_1(x),A_2(x))$ is $(\epsilon_1 + \epsilon_2, \delta_1+\delta_2)$-differentially private.



\begin{definition}
\label{def:sensitivity}
\textbf{Sensitivity of utility function~\cite{dwork:fttcs14}.} Given a dataset space $\D$ and an arbitrary range $\R$, a function $u: \D\times\R\to\mathbb{R}$  has a global sensitivity $\Delta_u$ defined as: $\Delta_u = \underset{r\in\R}\max \text{ }\underset{x\sim x'}\max |u(x,r) - u(x',r)|$
\end{definition}

\begin{definition}
\label{def:expo_mech}
\textbf{The Exponential Mechanism.} (Definition 3.4 of~\cite{dwork:fttcs14}) The exponential mechanism $M(x, u, \R)$ that outputs an element $r\in\R$ with probability proportional to $\exp(\frac{\epsilon u(x, r)}{\Delta_u})$ is $\epsilon$-differentially private, if the addition of an element to the data does not decrease the value of the utility function.
\end{definition}

\begin{theorem}
\label{theorem:expo-utility}
  \textbf{Utility of the Exponential Mechanism.} (Theorem 3.11 and Corollary 3.12 of~\cite{dwork:fttcs14}) For a given dataset $x$, let $\text{OPT}=max_{r\in\R}u(x, r)$. For the exponential mechanism $M(\cdot)$, we have:
\begin{center}
$\Pr\left[u(x, M(x, u, \R)) \leq \text{OPT} - \frac{2\Delta_u}{\epsilon}(\ln{|\R|}+r)\right] \leq e^{-r}$
\end{center}

\end{theorem}

\noindent 
\textbf{Adversarial Probabilistic Process.}
Following~\cite{gupta2010}, consider an adversarial probabilisic process with $n$ iterations. In each iteration $i$, the adversary tosses a coin, which gives heads with probability $p_i\in[0,1]$, adaptively by observing previous $i$-1  iterations. We utilize this process to prove the privacy bounds of our algorithms.

\begin{lemma}Lemma B.1 of~\cite{gupta2010}.
  \label{lemma:adversarial}
Let $Z_i$ be a random variable, with $Z_i= 1$ indicating that no head appears in the first $i$ iterations, and $Z_i=0$ otherwise. Let $Y=\sum_{i=1}^np_iZ_i$. For any $q$, $\Pr[Y>q] \leq \exp(-q)$.
\end{lemma}

\begin{corollary}
  \label{cor:adversarial}
  Let $\mathcal{T}$ be the first iteration in which a head appears in the process above. For any $\delta\in(0,1]$, $\sum_{i=1}^{\mathcal{T}-1}p_i \leq \ln{\delta^{-1}}$, with probability at least $1-\delta$.
\end{corollary}
  
\section{Related Works}
\label{sec:relatedworks}

Differential privacy has been a very active area of research since its introduction. We briefly summarize the main results and challenges; we focus only on works on privacy for graph that are directly relevant to our paper. Due to space constraints, we refer to~\cite{dwork2011differential, 10.1145/1065167.1065184, dwork:fttcs14, 10.1145/1250790.1250803, zhu:tkde17, Vadhan2017} for surveys and details of key results in this area, and provide additional discussion in the Appendix.

Much of the initial work on graph privacy involved computing different kinds of statistics, e.g., number of edges, counts of triangles and other subgraphs, and the cost of the minimum spanning tree
~\cite{10.1145/1250790.1250803, karwa:tds13, karwa2014differentially, mir2012differentially, hay2009accurate, Kasiviswanathan:2013:AGN:2450206.2450232, blocki:itcs13}. Graph statistics are challenging due to their high sensitivity.
\cite{10.1145/1250790.1250803} introduce a concept of \emph{smooth sensitivity} and apply it for triangle counting and minimum spanning tree problems in the edge privacy model.~\cite{karwa:tds13} generalize the method for other subgraph counting problems, such as k-star or k-triangle. Many other graph statistics in the edge-privacy model have been studied in~\cite{karwa2014differentially, mir2012differentially, hay2009accurate}. On the other hand, graph statistics in the node-privacy model are generally more difficult.~\cite{Kasiviswanathan:2013:AGN:2450206.2450232, blocki:itcs13} are two of the first studies that utilize Lipschitz Extensions to develop node differentially private mechanisms. They transform subgraph counting problems into linear programming optimizations of which the sensitivities of the solutions are restricted. Algorithms for node differentially private graph queries such as degree distributions or random graph model estimations have been established by~\cite{chen2013recursive, ding:cikm18, day2016publishing}.

In contrast to graph statistics, releasing a subset of nodes in the input graph with differential privacy (as in the st-mincut, $k$-median, or the vertex cover problems), or a collection of subsets in other combinatorial optimization problems (as in the set cover problem) is much more challenging, e.g.,~\cite{pmlr-v70-mitrovic17a,gupta2010}. 
One of the first papers on this was by~\cite{gupta2010}, who design private algorithms for different combinatorial optimization problems, including computing global mincuts, vertex cover and set cover. Our algorithms are inspired by the techniques of~\cite{gupta2010}.

Another direction of work is on generation of differentially private synthetic graphs~\cite{xiao2014differentially, nguyen2015differentially, chen2014correlated, gupta2012iterative}, which preserve some specific graph properties, such as cut, shortest path length and degree distribution queries. Such synthetic graphs are useful since post-processing does not cause any privacy violation. Finally, we note that private graph algorithms have also been considered for other notions of privacy, e.g.~\cite{imola2020locally}.

\label{sec:methods}

\section{Private Densest Subgraph}
\label{sec:sequential}
Here we design and analyze \seqalgo{} for the \prob{} problem. This is an adaptation of the (non private) algorithm Charikar~\cite{charikar:dense}, which gives a $1/2$-approximation for the densest subgraph problem---this algorithm constructs a sequence of subgraphs $S_1,\ldots,S_n$ (by removing a minimum degree node each time), and then chooses $\text{argmax}_{S_i} \rho(S_i)$. \seqalgo{} adapts this algorithm, and involves two parts.
First, it generates a sequence of candidate subgraphs by removing one node at a time using an Exponential Mechanism, with probability proportional to an exponential function of its current degree times a constant. Therefore, lower-degree nodes have exponentially more chances to be removed than higher-degree nodes. Intuitively, we try to remove low-degree nodes and retain high-degree nodes to construct dense candidates. The second part of the algorithm applies the Exponential Mechanism to randomly select one of the candidates by their densities.

\begin{algorithm}
  \caption{$\seqalgo(G, \epsilon, \delta)$\\
    \textbf{Input:} $G\in\Graphs$, privacy parameters $\epsilon, \delta$.\\
    \textbf{Output:} Subset $S\subseteq V(G)$ 
  }
  \label{alg:private-densest}
  \begin{footnotesize}
  \begin{algorithmic}[1]
    
    

    \STATE Let $\epsilon' = \frac{\epsilon}{4\ln{(e/\delta)}}$, $S_0 = V(G)$
    
    \FOR {$t$ $:=$ $1$ to $n$}
    
    \STATE Pick a node $\pi_t\in\{v: v\in S_{t-1}\}$ with probability proportional to $e^{-\epsilon' \deg_{S_{t-1}}(v)}$
    
    \STATE $S_t = S_{t-1} - \pi_t$
    
    \ENDFOR
    
    \STATE Return $S_t$ from $\{S_t: t=0,\ldots,n-1\}$ with probability proportional to $e^{\epsilon\rho(S_t)/2}$
  \end{algorithmic}
  \end{footnotesize}
\end{algorithm}

\begin{lemma}
  \label{lemma:sequential-exponential}
The computation of the sequence $\pi^G = \pi_1,\ldots,\pi_n$, in the \textbf{for} loop in Algorithm~\ref{alg:private-densest} (lines 3--5) is $(\epsilon/2, \delta)$-differentially private.
\end{lemma}
\begin{proof}
(Short, Full proof in Theorem~\ref{lemma:sequential-exponential-full})
Let $G\sim G'$ be two graphs that differ in exactly one edge $e=(u, v)$. Let $\pi^G$ and $\pi^{G'}$ denote the permutations computed in the for loop in Algorithm~\ref{alg:private-densest}. Let $\mathcal{T}$ is the first iteration that one of the endpoints of $e$ (i.e., $u$ or $v$) is removed. Let $adj(e)$ define the set of nodes adjacent to edge $e$.
Recall that $\degree{G}{j}$ is the degree of node $j$ in subgraph $S_t$ of graph $G$. Fix a permutation $\pi$. Consider the ratio:

{\centering
\scalebox{0.87}{\allowdisplaybreaks
$\begin{aligned}
  P &= \frac{\Pr[\pi^G = \pi]}{\Pr[\pi^{G'} = \pi]} \\
    &= \prod_{t=1}^{n} \frac{\exp(-\epsilon'\degree{G}{\pi_t})/\sum_{j}\exp(-\epsilon'\degree{G}{j})}{\exp(-\epsilon'\degree{G'}{\pi_t})/\sum_{j}\exp(-\epsilon'\degree{G'}{j})} \\
    &= \frac{\exp(-\epsilon'\text{deg}_{S_\T}^G(\pi_\T))}{\exp(-\epsilon'\text{deg}_{S_\T}^{G'}(\pi_\T))} \prod_{t=1}^\T\frac{\sum_j \exp(-\epsilon'\degree{G'}{j})}{\sum_j \exp(-\epsilon'\degree{G}{j})} 
\end{aligned}$}

The last equality follows from the definition of $\T$}, because: (i) for $t=\T + 1 ..n$, $\degree{G}{j}=\degree{G'}{j}$ for all nodes $j$, and, (ii) for $t = 1..\T-1$, $\degree{G}{\pi_t} = \degree{G'}{\pi_t}$. We have two cases for the analysis.

\textit{First}, suppose $G + e = G'$. In this case, $\frac{\exp(-\epsilon'\text{deg}_{S_\T}^G(\pi_\T))}{\exp(-\epsilon'\text{deg}_{S_\T}^{G'}(\pi_\T))} = \exp(\epsilon')$ because $\text{deg}_{S_\T}^G(\pi_\T) + 1 = \text{deg}_{S_\T}^{G'}(\pi_\T)$. Further, for each $t, j$, we have  $\degree{G'}{j} \geq \degree{G}{j}$ for all $j$, which implies the product $\prod_{t=1}^\T\frac{\sum_j \exp(-\epsilon'\degree{G'}{j})}{\sum_j \exp(-\epsilon'\degree{G}{j})}\leq 1$. Therefore, $P \leq \exp(\epsilon')$

\textit{Second}, suppose $G - e = G'$. Similar to the first case, $\frac{\exp(-\epsilon'\text{deg}_{S_\T}^G(\pi_\T))}{\exp(-\epsilon'\text{deg}_{S_\T}^{G'}(\pi_\T))} = \exp(-\epsilon') \leq 1$. 
For $\jAdjE, t\leq \T$, we have $\text{deg}_{S_t}^{G}(j) = \text{deg}_{S_t}^{G'}(j)+1$. For $\jnotAdjE$, and for all $t$, $\degree{G}{j} = \degree{G'}{j}$. Further, for $t>\T$, we have $\degree{G}{j} = \degree{G'}{j}$ for all $j$. Note that whenever we use $\degree{G}{j}$, we only consider $j\in S_{t-1}$, i.e., $j$ would not have been deleted from the graph before step $t$.\\
Therefore, for $t\leq\T$: we can expand the term $\sum_j\exp(-\epsilon'\degree{G'}{j})$ as $\sum_j\exp(-\epsilon'\degree{G'}{j}) = (e^{\epsilon'} - 1)\sum_\jAdjE\exp(-\epsilon'\degree{G}{j}) + \sum_j \exp(-\epsilon'\degree{G}{j})$. 
%
Substituting it in the expression for $P$. Since $\degree{G}{j} = \degree{G'}{j}$ for all $j$ and $t>\T$, we have

{\centering
{\scalebox{0.87}
  {\allowdisplaybreaks
    $\begin{aligned}
      P 
      &\leq \prod_{t = 1}^T (1 + (\exp(\epsilon') - 1)\frac{\sum_\jAdjE\exp(-\epsilon'\degree{G}{j})}{\sum_j\exp(-\epsilon'\degree{G}{j})}) \\
      &=\prod_{t = 1}^T (1 + (\exp(\epsilon') - 1)p_t(G))\\
      &\leq \prod_{t=1}^{\T}\exp((\exp(\epsilon')-1)p_t(G)),\mbox{ using $1+x\leq e^x$ for $x\geq 0$},
    \end{aligned}$}}

where $p_t(G)$ is the probability that an endpoint of $e$ is picked in iteration $t$.} We will show in Lemma~\ref{lemma:good}, by using the adversarial process (Section~\ref{sec:dp-basic}), that 
$\sum_{t=1}^{\T-1}p_t(G) \leq \ln\delta^{-1}$ with prob. at least $1 - \delta$ and use it to derive a bound on $P$. We say that $\pi$ is $\ln{\delta^{-1}}$-good if $\sum_{t=1}^{\T-1}p_t(G) \leq \ln\delta^{-1}$ and $\ln{\delta^{-1}}$-bad otherwise (\textit{good} and \textit{bad} for short). Therefore, with probability at least $1-\delta$, $\pi$ is \textit{good} and

{\centering
{\scalebox{0.82}
{
$\begin{aligned}
  P &\leq \prod_{t=1}^{\T}\exp((\exp(\epsilon')-1)p_t(G)) \\
    &\leq \exp(2\epsilon'\sum_{t=1}^\T p_t(G)), \mbox{ using $e^x\leq 1+2x$ for $x\leq 1$} \\
\end{aligned}$}}

}

{\centering
{\scalebox{0.88}
{
$\begin{aligned}
    &\leq \exp(2\epsilon'(\ln\delta^{-1} + p_\T(G)))
    \leq \exp(2\epsilon'(\ln\delta^{-1}+1)) \\
    &= \exp(2\epsilon'(\ln(e/\delta)) 
    = \exp(\epsilon/2), \mbox{ since $\epsilon' = \frac{\epsilon}{4\ln{(e/\delta)}}$},
\end{aligned}$}}

which implies $\Pr[\pi^G = \pi] \leq exp(\epsilon/2)\Pr[\pi^{G'}=\pi]$.}

Let $\mathcal{P}$ be the set of output orderings that might be generated by the \textbf{for} loop in \seqalgo. Then, we have

{\scalebox{0.88}
{\allowdisplaybreaks
$\begin{aligned}
  \Pr[\pi^{G} \in \mathcal{P}] &= \sum_{\pi\in\mathcal{P}}\Pr[\pi^{G} = \pi]\\
    &= \sum_{\substack{\pi\in\mathcal{P}:\text{$\pi$is \textit{good}}}}\Pr[\pi^{G} = \pi]
    + \sum_{\substack{\pi\in\mathcal{P}:\text{$\pi$ is \textit{bad}}}}\Pr[\pi^{G} = \pi] \\
  &\leq\exp(\epsilon/2)\Pr[\pi^{G'}\in\mathcal{P}] + \delta
\end{aligned}$}}

The lemma then follows.
\end{proof}

\begin{lemma}
  \label{lemma:good}
  Let $p_t(G)$ is the probability that an endpoint of edge $e$ is pick in iteration $t$ and $\T$ is the first iteration that an endpoint of edge $e$ is picked (as stated in Lemma~\ref{lemma:sequential-exponential}), $\sum_{t=1}^{\T-1}p_t(G) \leq \ln{\delta^{-1}}$ with probability at least $1-\delta$.
\end{lemma}

\begin{proof}
We map our algorithm to the adversarial probabilistic process in Lemma~\ref{lemma:adversarial} and Corollary~\ref{cor:adversarial}. In each iteration $t$, the adversary chooses heads with probability equal to $p_t(G)$. Since $p_t(G)$ is calculated based on the outcomes of $t-1$ previous iterations, it satisfies the conditions of the adversarial model. In any iteration $t$, if no head has appeared in the previous $t-1$ iterations, we have $p_t(G)= \frac{\sum_{ \jAdjE}\exp(-\epsilon'\degree{G}{j})}{\sum_j \exp(-\epsilon'\degree{G}{j})}$, since each node $j$ is picked with probability proportional to $\exp(-\epsilon'\degree{G}{j})$.
Therefore, the two processes (i.e., our algorithm and the adversarial probabilistic process) are now equivalent. By Corollary~\ref{cor:adversarial}, $\sum_{t=1}^{\T-1}p_t(G) \leq \ln\delta^{-1}$ with probability at least $1 - \delta$. 
The lemma then follows.

\end{proof}

\begin{theorem}
  \label{theorem:sequential-alg}
 Algorithm~\ref{alg:private-densest} is $(\epsilon, \delta)$-differentially private.
\end{theorem}
\begin{proof}
By Lemma~\ref{lemma:sequential-exponential}, the sequence $\{S_1, S_2, ...\}$ is $(\epsilon/2, \delta)$-differentially private. For each $S_t$, adding or removing 1 edge to/from the graph changes $\rho(S_t)$ by at most 1, hence $\Delta_{\rho} = 1$. Applying the Exponential Mechanism, the last command releases an output with $\epsilon/2$-differential privacy. Using the Composition Theorem, Algorithm~\ref{alg:private-densest} is $(\epsilon, \delta)$-differentially private.
\end{proof}

\begin{lemma}
\label{lemma:seq-vt}
In each iteration $t$, the node $v_t$ picked by the exponential mechanism satisfies $deg_{S_t}(v) \leq 2\rho(S_t) + \frac{24}{\epsilon}\ln(e/\delta)\ln(n)$, with probability at least $1-1/n^2$.
\end{lemma}
\begin{proof}
Since $2\rho(S_t) = \frac{2(\mbox{\#edges in $S_t$)}}{|S_t|} = \frac{\sum_{v\in S_t} deg_{S_t}(v)}{|S_t|}$, it follows that $\min_{v\in S_t} deg_{S_t}(v)\leq 2\rho(S_t)$. Let $v_{min}$ be a node achieving the minimum degree in $S_t$.
Then $deg_{S_t}(v_{min}) \leq 2\rho(S_t)$. \\
In the exponential mechanism run in iteration $t$, the sensitivity $\Delta=1$, as $|deg^G_{S_t}(v) - deg^{G'}_{S_t}(v)|\leq 1$. Applying Theorem~\ref{theorem:expo-utility} (with $u(S_t, v) = -\degree{G}{v}$ and $r = 2\ln{n}$), it follows that $\Pr[deg_{S_t}(v_t) \geq deg_{S_t}(v_{min}) + \frac{6}{\epsilon'}\ln(n)]\leq e^{-2\ln{n}}\leq 1/n^2$. Substituting $\epsilon' = \frac{\epsilon}{4\ln(e/\delta)}$, the Lemma follows.
\end{proof}

\begin{theorem}
\label{theorem:approx-seq}
Let $S^*$ be an optimal solution. Assuming $n > 3, \delta < 1/e$, the set $S$ output by Algorithm~\ref{alg:private-densest} satisfies $\rho(S)\geq \frac{\rho(S^*)}{2} - \frac{32}{\epsilon}\ln(1/\delta)\ln(n)$, with probability at least $1-2/n$.
\end{theorem}
\begin{proof}
Applying a union bound with Lemma~\ref{lemma:seq-vt}, it follows that for each iteration $t$, $deg_{S_t}(v_t)\leq 2\rho(S_t) + \frac{24}{\epsilon}\ln(e/\delta)\ln(n)$, with probability $1-1/n$.
Let $T$ denote the first iteration that a node $v_T\in S^*$ is picked. As observed in~\cite{bahmani2012densest}, $deg_{S^*}(v) \geq \rho(S^*)$, therefore

{\centering
{\scalebox{0.88}
{
$\begin{aligned}
  \rho(S^*) &\leq deg_{S^*}(v) \leq deg_{S_T}(v) \\
            &\leq 2\rho(S_T) + \frac{24}{\epsilon}\ln(e/\delta)\ln(n),
\end{aligned}$}}

with probability at least $1-1/n$.}
Therefore, $\rho(S_T) \geq \rho(S^*)/2 - \frac{12}{\epsilon}\ln(e/\delta)\ln(n)$, with prob. at least $1-1/n$.

Next, applying Theorem~\ref{theorem:expo-utility} to the exponential mechanism in the last line of the algorithm, we have $\Pr[\rho(S) \leq \rho(S_T) - 4\frac{\log{n} + r}{\epsilon}] \leq e^{-r}$. Choosing $r=\ln{n}$, we have $\rho(S)\geq \rho(S_T)- 8\ln{n}/\epsilon$, with probability at least $1-1/n$. Combining this with the bound on $\rho(S_T)$, we have $\rho(S)\geq \rho(S^*)/2 - \frac{12}{\epsilon}\ln(e/\delta)\ln(n) - 8\ln{n}/\epsilon$. 

Finally, we derive a coarse but simpler bound. For $\delta\leq 1/e$, we have  $1\leq\ln(e/\delta)\leq 2\ln(1/\delta)$. Using these in the lower bound for $\rho(S)$ gives $\rho(S)\geq
\frac{\rho(S^*)}{2} - ({24} +\frac{8}{ln(1/\delta)})\frac{1}{\epsilon}\ln(1/\delta)\ln(n)\geq
\frac{\rho(S^*)}{2} - \frac{32}{\epsilon}\ln(1/\delta)\ln(n)$.
\end{proof}

\section{Parallel Private Densest Subgraph}


\subsection{Algorithm \palalgo{}}

Instead of picking a single node in each iteration, as in Algorithm \seqalgo{}, the main idea in \palalgo{} is to independently sample each node with the right probability. As a result, multiple nodes might be selected simultaneously, especially when the degrees are low.
This change makes the privacy analysis more challenging because (1) we have to consider different scenarios of picking the endpoints of the extra edge $e$ and (2) without the normalization terms, the probability $P$ in the proof of Lemma~\ref{lemma:sequential-exponential} loses its symmetry.


\begin{algorithm}
  \caption{$\palalgo{}(G, \epsilon, \delta)$\\
    \textbf{Input:} $G\in\Graphs$, privacy parameters $\epsilon, \delta$.\\
    \textbf{Output:} A subset $S\subset V(G)$ 
  }
  \label{algo:parallel}
  \begin{footnotesize}
  \begin{algorithmic}[1]
    
    

    \STATE Let $\epsilon' = \frac{1-1/e}{8\ln{(e/\delta)}}\epsilon$, and $c = 1/\epsilon' + 1$

    \STATE $S_0 = V(G)$, $t = 0$
    
    \WHILE {$|S_t| > 0$}
    
    \STATE $\forall j \in S_t$, assign a random variable $p_{tj} \in \{0, 1\}$ such that $\Pr[p_{tj}=1] = P_{tj} = \exp(-\epsilon'(\text{deg}_{S_t}(j) + c))$
    
    \STATE $t := t + 1$

    \STATE $\pi_t = \{j \in S_t: p_{tj}=1\}$
    
    \STATE $S_t = S_{t-1} \backslash \pi_t$
    
    \ENDWHILE
    
    \STATE Return $S_t$ from distinct set $\{S_t\}$ with probability proportional to $e^{\epsilon\rho(S_t)/2}$
  \end{algorithmic}
  \end{footnotesize}
\end{algorithm}

\begin{theorem}
  \label{theorem:parallel-alg}
  Algorithm~\ref{algo:parallel} is $(\epsilon, \delta)$-differentially private.
\end{theorem}
 \begin{proof}
(Sketch, full proof in Theorem~\ref{theorem:parallel-alg-full})
Using the same notations in Lemma~\ref{lemma:sequential-exponential}, we prove the sequence $\{S_1, S_2, ...\}$ output by the \textbf{while} loop is $\left(\epsilon/2, \delta\right)$-differentially private. The remaining part is the same to Theorem~\ref{theorem:sequential-alg}. 
We construct $P = \frac{\Pr[\pi^G = \pi]}{\Pr[\pi^{G'} = \pi]}$, and define:

{\centering
{\scalebox{0.88}
{
$\begin{aligned}
A_t = \prod_{j: p_{tj}=1}\frac{P_{tj}^G}{P_{tj}^{G'}}&, 
B_t = \prod_{j: p_{tj}=0}\frac{1-P_{tj}^G}{1-P_{tj}^{G'}},\\
  C = \prod_{j: p_{\T j}=1}\frac{P_{\T j}^G}{P_{\T j}^{G'}}&,
  D = \prod_{j: p_{\T j}=0}\frac{1-P_{\T j}^G}{1-P_{\T j}^{G'}}
\end{aligned}$}},

then expand} $P = \prod_{t=1}^{\T-1}(A_t \times B_t) \times C \times D$.

We consider 2 cases (1) $G + e = G'$ and (2) $G - e = G'$ and 2 subcases in each case: (a) both $u, v$ are picked at step $\T$ and (b) exact one of them is picked at $\T$. In Case 1, it is straightforward to prove $P \leq \exp(2\epsilon')$, where Case 2 is more difficult to analyze.

In Case 2, we prove $A_t=1 \forall t < \T$. We reduce $C\text{ and } D$, and depends on the subcases, they have slightly different forms. We then reduce $P$ to (2a) $P \leq \prod_{t =1}^{\T -1}B_t$ and (2b) $P \leq \prod_{t =1}^{\T -1}B_t \times  \frac{1-P_{\T u}^G}{1-P_{\T u}^{G'}}$. Hence, the remaining thing is to bound $\prod_{t =1}^{\T -1}B_t$. Since for all $j\not\in \text{adj}(e)$, $\degree{G}{j} = \degree{G'}{j}$, the terms corresponding to such nodes cancel out in $B_t$, which leads to $\prod_{t=1}^{\T-1}B_t \leq \exp(\frac{4\epsilon'}{1-e^{-1}}\sum_{t=1}^{\T-1}p^{uv}_t(G))$, in which $p^{uv}_t(G)$ is the probability of picking at least one of $u, v$ in iteration $t$. Mapping the Algorithm's sampling process to the Adversarial Probabilistic Process in Lemma~\ref{lemma:adversarial}, we find the upper bound of $\sum_{t=1}^{\T-1}p^{uv}_t(G)$ is $\ln{\delta^{-1}}$ with probability at least $1-\delta$. It follows that $P \leq \exp(\frac{4\epsilon'\ln(1/\delta)}{1-1/e})\leq\exp(\epsilon/2)$ in Case 2a and $P \leq \exp(\frac{4\epsilon'\ln(e/\delta)}{1-1/e}) = \exp(\epsilon/2)$ in Case 2b. Using the same argument as in the last part of Lemma~\ref{lemma:sequential-exponential}, the proof follows. 
 \end{proof}

\begin{theorem}
\label{theorem:parallel-accuracy}
(Proof in Theorem~\ref{theorem:parallel-accuracy-full}) 
Let $S^*$ be an optimal solution. If $\delta\leq 1/e$,
the set $S$ output by Algorithm~\ref{algo:parallel} satisfies $\rho(S)\geq \frac{\rho(S^*)}{2} - \frac{56}{\epsilon}\ln(1/\delta)\ln{n}$, with probability at least $1-2/n$.
\end{theorem}
%

\noindent
\textbf{Running time of \palalgo{}.}  Each node is removed in an iteration with probability $e^{-\epsilon' \deg_{S_{t-1}}(v)}$. This could be as small as $e^{-\Theta(n)}$, e.g., if $S_{t-1}$ has a dense subgraph in which each node has very high degree. Therefore, the number of iterations could be $e^{\Theta(n)}$ in the worst case (this is borne out of our experiments on some networks). 


\subsection{Algorithm \phasealgo{}: Logarithmic Bounded Runtime Algorithm}
Algorithm~\ref{algo:parallel} can execute in parallel but it has exponential runtime. This section presents a modified version of \palalgo{}~that has logarithmic bounded runtime, named~\phasealgo{}. We group iterations into phases and aim to remove at least a constant fraction of nodes in each phase (Lemma~\ref{lemma:phase-iterations}). Also, we only update node's new degrees at the end of each phase, allowing the sampling process per each node to run independently as its degree remains unchanged in the current phase.
The privacy analysis of~\phasealgo{} differs from \palalgo{}'s in two key ways: (1) the terms corresponding to the endpoints of the extra edge $e$ are not updated until the end of the current phase and (2) it requires two instances of the Adversarial Probabilistic Process to mimic the sampling process.
\begin{algorithm}
  \caption{\phasealgo{}$(G, \epsilon, \delta)$\\
    \textbf{Input:} $G\in\Graphs$, privacy parameters $\epsilon, \delta$.\\
    \textbf{Output:} A subset $S \subseteq V(G)$
  }
  \label{algo:parallel-phases}
  \begin{footnotesize}
  \begin{algorithmic}[1]
    
    \STATE Let $\epsilon' = \frac{1-1/e}{24\ln{(4/\delta)}}\epsilon$, and $c = 1/\epsilon' + 1$
    \STATE Let $\pi_0 = V, S_0 = V$, $T_0 = 0$, $i=1$

    \WHILE{$S_i\neq\emptyset$}
    \IF{$|S_i|\leq\ln{n}$}
    \STATE $T_{i} = T_{i-1}+1$, $\pi_{T_{i}} = S_i$ $S_{i+1}=\emptyset$, $i=i+1$
    \ELSE
    \STATE For each $v\in S_i$, let $d(v) = deg_{S_i}(v)$
    \STATE Let $P_{v} = \exp(-\epsilon'(d(v)+c))$
    \STATE Let $T_v$ be chosen from a geometric random process with probability $P_v$
    \STATE Let $\widehat{\rho(S_i)}=\rho(S_i)+\frac{16}{\epsilon}\ln{n}+Lap(\frac{4\ln{n}}{|S_i|\epsilon})$
    \STATE Let $T_i=exp(\epsilon'(4\widehat{\rho(S_i)} + c))\cdot 4\ln{n}$ 
    \STATE For $t\leq T_i$, let $\pi_t=\{v: T_v = t\}$
    \STATE $S_{i+1} = S_i \setminus \cup_{t'={T_{i-1}}+1}^{T_i} \pi_{t'}$
    \STATE $i = i+1$
    \ENDIF
    \ENDWHILE
    
    \STATE Return $S\in \text{ distinct set }\{S_0, S_1,\ldots,\}$ with probability proportional to $e^{\epsilon\rho(S_i)/2}$
  \end{algorithmic}
  \end{footnotesize}
\end{algorithm}

\begin{lemma}
  \label{lemma:phase-iterations-full}
  \label{lemma:phase-iterations}
The number of phases is at most $\log{n}$, with probability at least $1-\frac{1}{n^2}$.
\end{lemma}
\begin{proof}
Consider any phase $i$, except the last one, and let $v\in S_i$ be a node with $d_{S_i}(v)\leq 4\rho(S_i)$. Since $T_v$ is sampled from a geometric distribution, $\Pr[T_v > 4\ln{n}/P_v]\leq (1-P_v)^{4\ln{n}/P_v}\leq 1/n^4$. Therefore, with probability at least $1-1/n^3$, for all nodes $v\in S_i$, we have $T_v\leq exp(\epsilon'(d_{S_i}(v)+c))4\ln{n}\leq exp(\epsilon'(4\rho(S_i)+c))4\ln{n} \leq exp(\epsilon'(4\widehat{\rho(S_i)}+c))4\ln{n}= T_i$, where the second inequality is because $d_{S_i}(v)\leq 4\rho(S_i)$ and the last inequality follows Lemma~\ref{lemma:phase-density}. Let $A(S_i)=\cup_{t'=T_{i-1}+1}^{T_i} \pi_{t'}$ denote the set of nodes removed in this phase.

Therefore, with probability at least $1-1/n^3$, every node $v\in S_{i+1}=S_i-A(S_i)$ has $d_{S_i}(v) > 4\rho(S_i)$. We have
$2|E(S_i)| = \sum_{v\in A(S_i)} d_{S_i}(v) + \sum_{v\in S_{i+1}} d_{S_i}(v)\geq |S_{i+1}| 4\rho(S_i) = 4|S_{i+1}|\frac{|E(S_i)|}{|S_i|}$, which implies $|S_{i+1}|\leq |S_i|/2$. Thus, with probability at least $1-1/n^3$, the number of nodes in each phase reduces by factor of 2. Therefore, with probability at least $1-\frac{\log{n}}{n^3}\geq 1-\frac{1}{n^2}$, the Lemma follows.
\end{proof}

\begin{theorem}
  \label{theorem:parallel-phases-privacy}
Algorithm~\ref{algo:parallel-phases} preserves $(\epsilon, \delta)$-differential privacy, with $\epsilon\in(0,1], \delta\in(2n^{-2},1)$.
\end{theorem}

%

\begin{proof} (Sketch; full proof in Theorem~\ref{theorem:parallel-phases-privacy-full})
Similar to Theorem~\ref{theorem:parallel-alg}, we prove the sequence $\{S_1, S_2, ...\}$ output by the \textbf{while} loop is $(\epsilon/2, \delta)$-differentially private.

\emph{First,} prove the sequence $\{\pi_0, \pi_1, \ldots\}$ is $(\epsilon/4, \delta/2)$-differentially private. Using the same notations and expansion, we re-analyze the same cases in Theorem~\ref{theorem:parallel-alg}. Analysis in Case 1a and 2a (where $u$ and $v$ are removed at the same iteration) remains unchanged. We re-evaluate Case 1b and 2b. Assume $u$ is removed first at $\T_u$ and $v$ is removed later at $\T = \T_v > \T_u$.

In Case 1b, we prove that $A_t=1\forall t \neq \T_u$ and $A_{\T_u}=exp(\epsilon')$, $C$ and $D$ are the same as in Theorem~\ref{theorem:parallel-alg}. We split the product of $B_t$ into 2 ranges $B_t = \frac{(1-P_{tu}^G)(1-P_{tv}^{G})}{(1-P_{tu}^{G'})(1-P_{tv}^{G'})}$ for $t \in [1, \T_u-1]$ and $B_t = \frac{1-P_{tv}^{G}}{1-P_{tv}^{G'}}$ for $t \in [\T_u, \T_v-1]$. In either case, $B_t\leq 1$. We have $P= \prod_{t = 1}^{\T_v-1}(A_t\times B_t)\times C \times D \leq \exp(2\epsilon')$.

In Case 2b, similar to Theorem~\ref{theorem:parallel-alg}, we prove $A_t \leq 1, C \leq 1, D = 1$, and split the product of $B_t$ into two parts, where the latter arises since we do not update the degree of $v$ after removing $u$: $\prod_{t=1}^{\T_v-1}B_t = \prod_{t=1}^{\T_u-1}\frac{(1-P_{tu}^G)(1-P_{tv}^{G})}{(1-P_{tu}^{G'})(1-P_{tv}^{G'})}  \times \prod_{t=\T_u}^{\T_v-1}\frac{1-P_{tv}^{G}}{1-P_{tv}^{G'}} \leq\prod_{t=1}^{\T_u-1}\exp(\frac{4\epsilon'p_t(G)}{1-e^{-1}}) \times \prod_{t'=1}^{\T_v-\T_u-1}\exp(\frac{2\epsilon'q^v_{t'}(G)}{1-e^{-1}})$.
We use two instances of the Adversarial Probabilistic Process to calculate the upper bound on the product of $B_t$ as $\prod_{t=1}^{\T_v-1}B_t \leq \exp(\frac{4\epsilon'\ln{(4/\delta)}}{1-e^{-1}})\times \exp(\frac{2\epsilon'\ln{(4/\delta)}}{1-e^{-1}}) = \exp(\frac{6\epsilon'\ln{(4/\delta)}}{1-e^{-1}})$, with probability at least $1-\delta/2$.\\
\emph{Second}, we prove the sequence $\{\widehat{\rho(S_0)}, \widehat{\rho(S_0)}, \ldots\}$ is $(\epsilon/4, \delta/2)$-differentially private, using the Laplace mechanism on each $\widehat{\rho(S_i)}$ and composition over $\ln{n}$ elements of the sequence. \emph{Finally}, Because $\{S_0, S_1,\ldots\}$ is composed from $\{\pi_0, \pi_1,\ldots\}$ and the the information computed from $\{\widehat{\rho(S_0)}, \widehat{\rho(S_1)}, \ldots\}$, it is $(\epsilon/2,\delta)$-differentially private.
\end{proof}

\begin{theorem}
\label{theorem:parallel-phases-accuracy}
(Proof in Theorem~\ref{theorem:parallel-phases-accuracy-full})
Let $S^*$ be an optimal solution.
With probability at least $1-2/n$, the set $S$ output by Algorithm~\ref{algo:parallel-phases} satisfies $\rho(S)\geq \frac{\rho(S^*)}{4} -  \frac{160}{\epsilon}\ln(1/\delta)\ln(n)$ if $\delta \leq 1/e$.
\end{theorem}
%

\section{Map-Reduce Implementation}
\label{sec:mapr}
We briefly describe \mralgo{} (Algorithm~\ref{algo:mapr-parallel-phases})--a Map-Reduce implementation of \phasealgo{}. We prove in Lemma~\ref{lemma:map-reduce} that \mralgo{} preserves the same privacy guarantee and runtime as of \phasealgo.\\
Each iteration $i$ of \mralgo~contains 3 Reduce subphases and 1 Map subphase and is equivalent to phase $i$ of \phasealgo. The first Reduce subphase takes input as pairs of nodes that constitute an edge and group them by endpoints. It creates a list of neighbors of each node to keep track of current nodes and calculate their degrees. The second Reduce subphase groups the current nodes together to form a candidate set $S_i$ while the third one gathers the nodes' degrees to calculate the density $S_i$. The Map subphase performs the sampling process. It samples the removal time of each node and compares to the cut-off time of the current phase to decide if the node is picked or not. If the node is picked, it emits the node's self-edge to signal the next iteration's first Reduce subphase that the node is removed. Otherwise, it emits the edges adjacent to the node for the next iteration to rebuild the graph.

\section{Lower bound}
\label{sec:lb}

\begin{theorem}
Any $(\epsilon,\delta)$-differentially private algorithm for \prob{} must incur an additive lower bound of $\Omega(\sqrt{\log{n}/\epsilon})$ for $\delta \leq 1/n$.
\end{theorem}

\begin{proof}
Let $G=(V, E)$ denote a graph with $|V|=n$ and $E=\emptyset$. 
Let $M(G)$ be an $(\epsilon, \delta)$-differentially private mechanism.
Let $V$ be partitioned into sets $V_1,\ldots,V_N$, for $N=n/a$, ${a\choose 2}=\frac{1}{3\epsilon}\log{n}$, with each $|V_i|=a$. For $i\in\{1,\ldots,N\}$, define $\mathcal{S}_i = \{S\subseteq V: S\cap V_i\neq\emptyset, |S|\leq a^2\}$.
We prove in Lemma~\ref{lemma:lower-prob} that there exists an $i\in\{1,\ldots,N\}$ such that $\Pr[M(G)\in\mathcal{S}_i] \leq \frac{2a^3}{n}$.

Next, let $G'=(V, E')$ be a graph such that $V_i$ is a clique in $E'$, and there are no additional edges. Then, $|E'|={a\choose 2} = \frac{1}{3\epsilon}\log{n}$.
By the group privacy property of $(\epsilon, \delta)$-differential privacy (Lemma 2.2 of~\cite{Vadhan2017}), it follows that $\Pr[M(G')\in\mathcal{S}_i] \leq \Pr[M(G)\in\mathcal{S}_i] \exp(\frac{1}{3}\log{n}) + \frac{1}{3\epsilon}\log{n}\exp(\frac{1}{3}\log{n})\delta \leq \frac{2a^3}{n}n^{1/3} + \frac{\log{n}}{3\epsilon}n^{1/3}\frac{1}{n} = \frac{2a^3}{n^{2/3}} + \frac{\log{n}}{3\epsilon n^{2/3}} \leq \frac{3a^3}{n^{2/3}}$.




For any $S\not\in \mathcal{S}_i$, we have $\rho(S)\leq 1/2$, since either (1) $|S|>a^2$, which implies $\rho(S)\leq \frac{{a\choose 2}}{a^2} \leq \frac{1}{2}$, since $S$ may have up to ${a\choose 2}$ edges.
or (2) $|S|\leq a^2, S\cap V_i=\emptyset$, which implies $\rho(S)=0$.
If $M(G')\in\mathcal{S}_i$, we have $\rho(M(G'))\leq a/2$.
Hence, $\mathbb{E}[\rho(M(G'))]\leq \frac{a}{2}\Pr[M(G')\in\mathcal{S}_i] + \frac{1}{2} \cdot (1-\Pr[M(G')\in\mathcal{S}_i])\leq \frac{3a^4}{2n^{2/3}} + \frac{1}{2} \leq 1$, and the theorem follows since $\rho(G')=\frac{a-1}{2}$.

\end{proof}
\section{Experiments}
\label{sec:experiments}
Our experiments address the following questions.\\
\textbf{Accuracy of our proposed methods.} How does the accuracy of the solution computed using our private algorithms vary with $\epsilon$ and $\delta$ in different graphs?\\
\textbf{Recall.} To what extent can our private algorithms find the densest subgraph, as quantified by the recall?\\
\textbf{Efficiency of the parallel algorithms.} 
How does the number of iterations vary, relative to \seqalgo{}?
\subsection{Baselines and measurements}
We use the algorithm of~\cite{charikar:dense} as a baseline to compute a \textit{non-private densest solution} for performance evaluation in our experiments. Let $S_b$ denote the subgraph output by the baseline algorithm, and let $\mathcal{M}(G)$ denote the solution computed by our private algorithm $\mathcal{M}$ on graph $G$ (for fixed $\epsilon, \delta$). We consider the following metrics for evaluating our algorithms.
For all these metrics, the closer they are to $1$, the better our algorithms perform, in comparison to the baseline.

$\bullet$ \emph{Relative density}: this is defined as $\rho(\mathcal{M}(G))/\rho(S_b)$. \\
$\bullet$ \emph{Jaccard index}: $|\mathcal{M}(G)\cap S_b|/|\mathcal{M}(G)\cup S_b|$, which captures similarity between $\mathcal{M}(G)$ and $S_b$. \\
$\bullet$ \emph{Recall}: $|\mathcal{M}(G)\cap S_b|/|S_b|$, which quantifies what fraction of nodes in $S_b$ are selected by $\mathcal{M}(G)$. 

Table~\ref{tab:nets} lists 6 different networks from SNAP database we use to evaluate our results---these are chosen to be of various sizes, in order to understand the impact of network structure on the results~\cite{snapnets}. 
Due to the space limit, we present the setup and experiments on an extended list of 20 networks (Table~\ref{tab:full-nets}) in Section~\ref{sec:setup}. Also, we only show results for \seqalgo{} and \palalgo{} here; see the Appendix for a detailed comparison with \phasealgo{}.
\begin{table}[]
\footnotesize
\begin{tabular}{llll}
\textbf{Network}        & \textbf{Nodes}  & \textbf{Edges}  & \textbf{Description}\\
ca-GrQc        & 5242   & 14496  & Collab. net. Arxiv General Rel.\\
musae\_DE      & 9498   & 153138 & Social net. of Twitch (DE)\\
musae\_ENGB    & 7126   & 35324  & Social net. of Twitch (ENGB)\\
ca-AstroPh     & 18772  & 198110 & Collab. net. Arxiv Astro Phy.\\
musae\_squirrel& 5201   & 198493 & Wiki page-page net. (squirrel)\\
facebook       & 4039   & 88234  & Social circles from Facebook
\end{tabular}
\caption{\textbf{Summary of 6 of the 20 networks used for the experiments}~\cite{snapnets}; the remaining networks are summarized in Table~\ref{tab:full-nets} of the Appendix.}
\label{tab:nets}
\end{table}

\subsection{Experimental Results}
\begin{figure}[h]
  \includegraphics[width=\columnwidth,keepaspectratio]{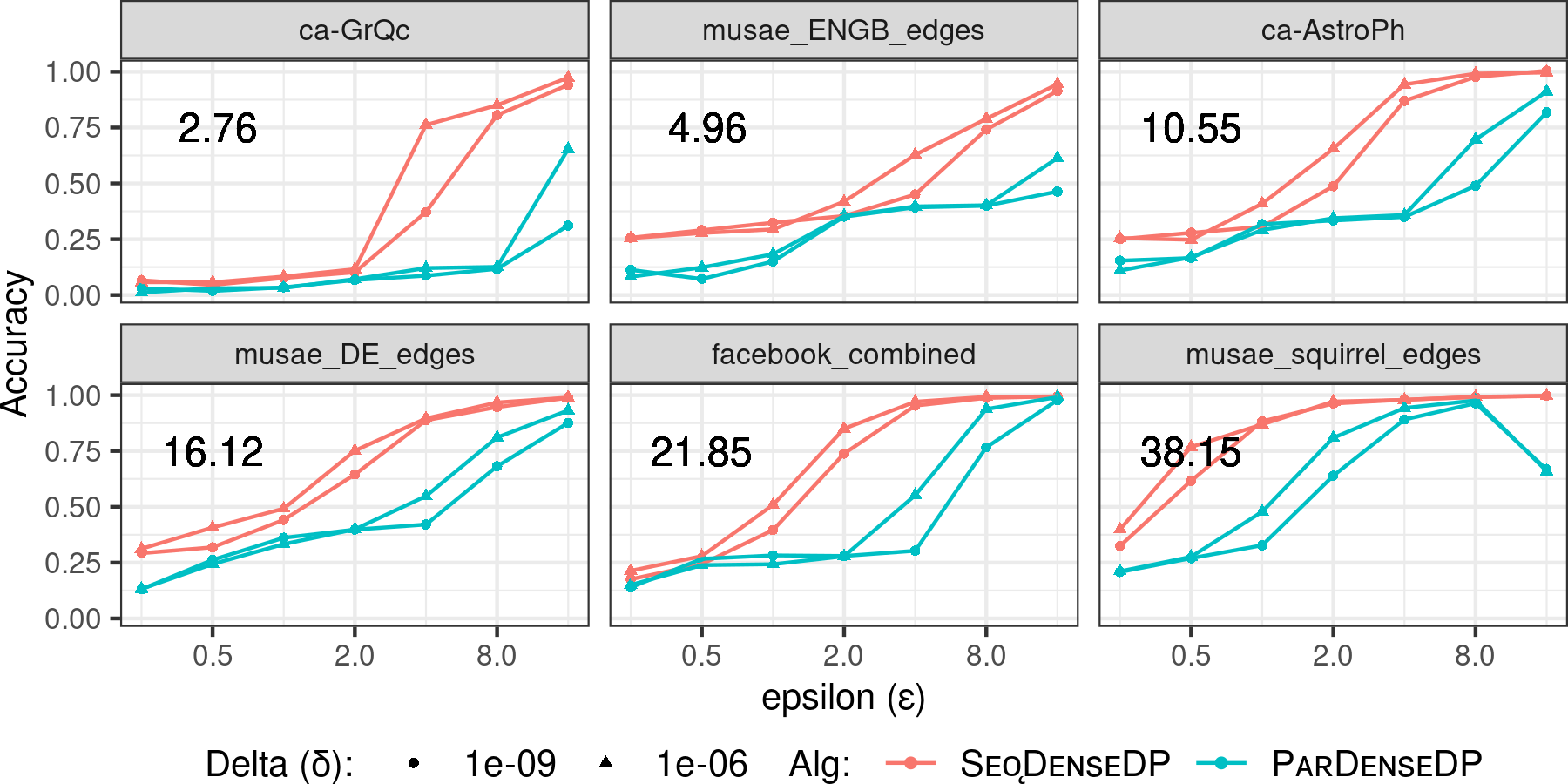}
  \centering
  \caption{\textbf{Accuracy in term of relative density of our private algorithms.}  The number in each graph indicates the density of the graph as whole (which equals the average degree).}
  \label{fig:accuracy}
\end{figure}

\begin{figure}[h]
  \includegraphics[width=\columnwidth,keepaspectratio]{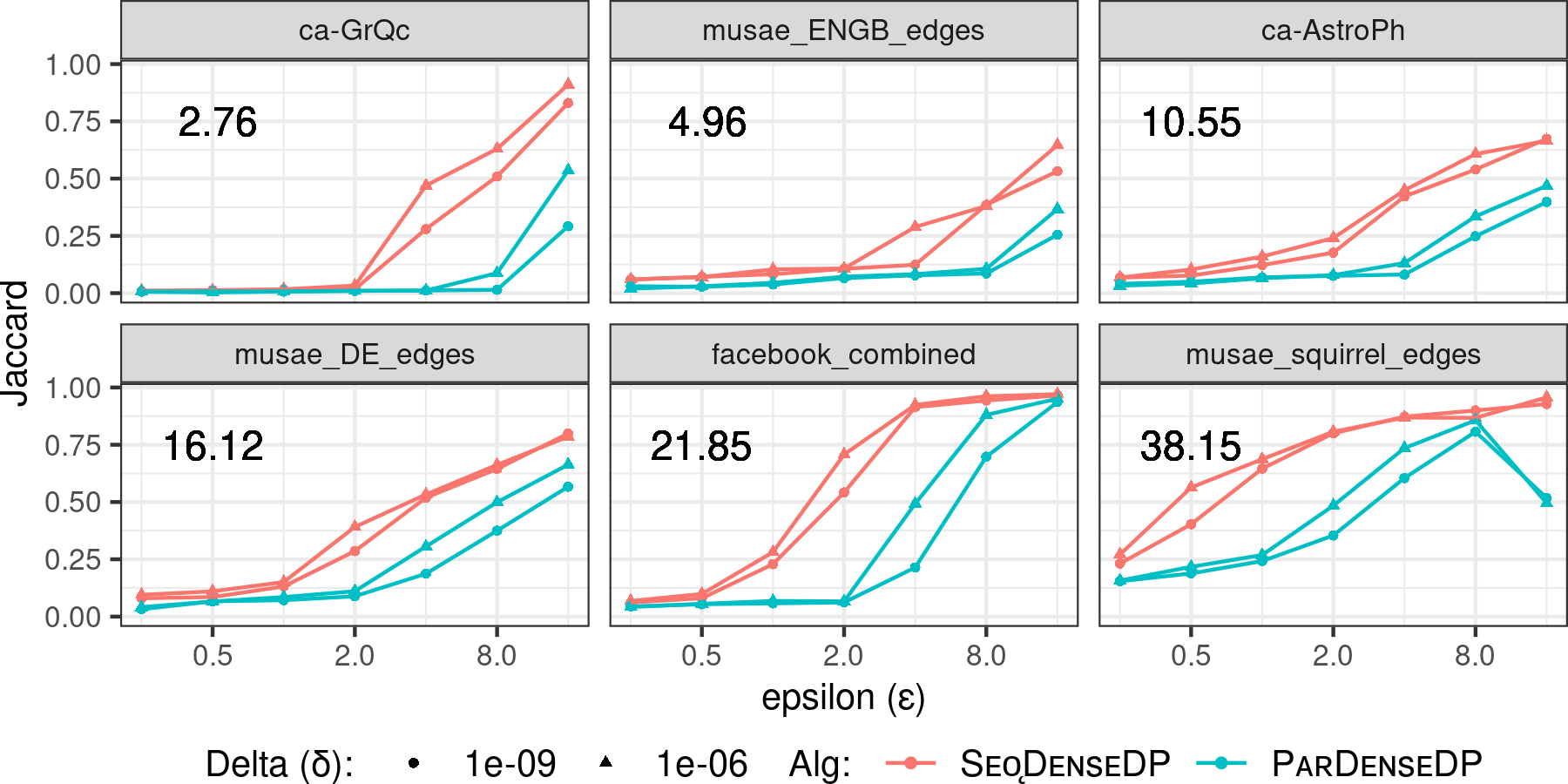}
  \centering
  \caption{\textbf{Jaccard similarity coefficient of private subgraphs and non-private baselines.}}
  \label{fig:jaccard}
\end{figure}

\begin{figure}[h]
  \includegraphics[width=\columnwidth,keepaspectratio]{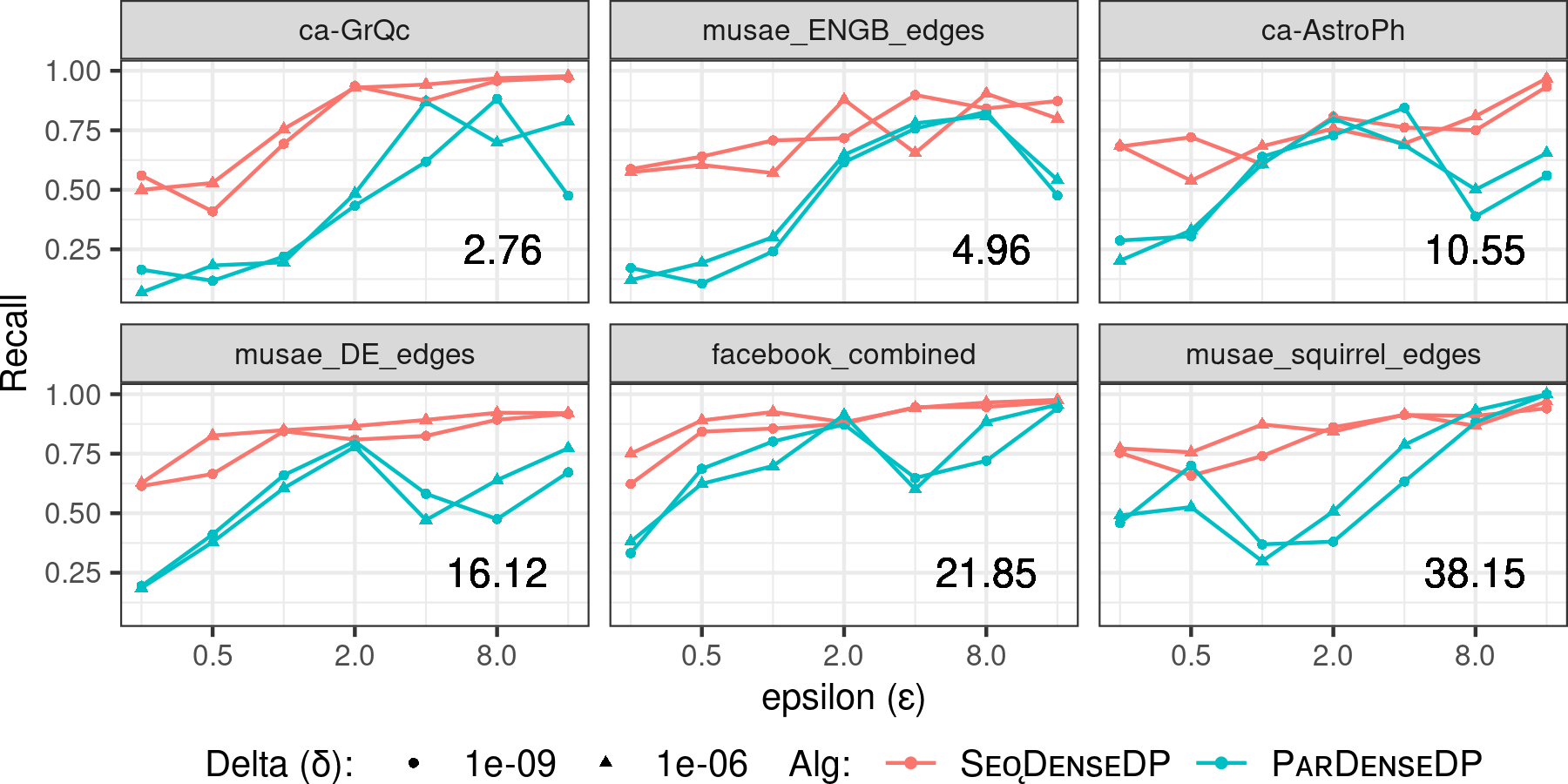}
  \centering
  \caption{\textbf{Recall: Fraction of nodes of the baseline's subgraphs are included in our algorithm's outputs.} }
  \label{fig:recall}
\end{figure}

\begin{figure}[h]
  \includegraphics[width=\columnwidth,keepaspectratio]{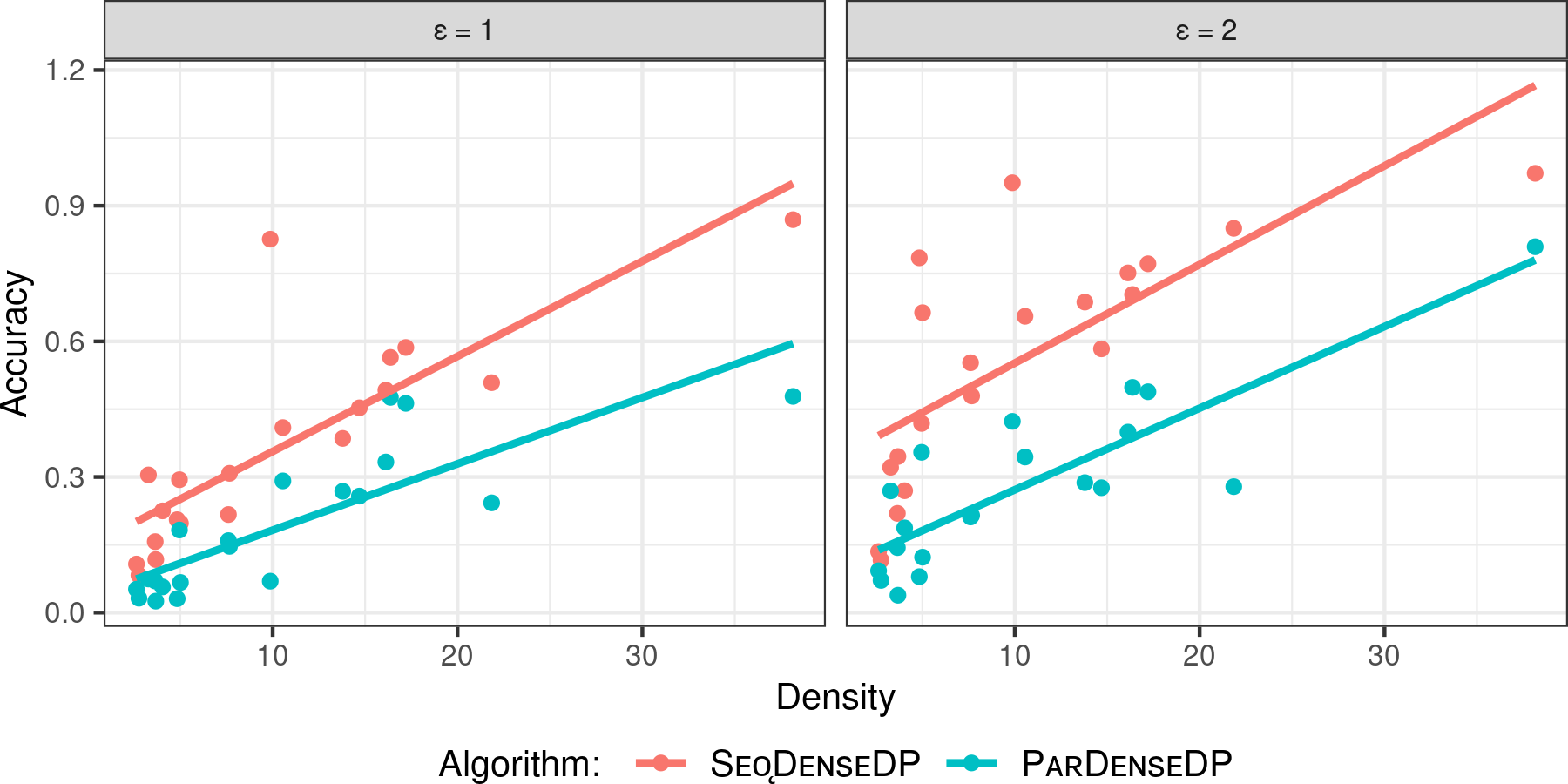}
  \centering
\caption{\textbf{The relationship between accuracy (relative density) and network density.}
We use $\epsilon\in\{1, 2\}, \delta=10^{-6}$. Each point corresponds to one network. The solid lines show linear models for the relative density and the network density. The plot confirms that our algorithms have better accuracy on higher density networks.}
\label{fig:densityVsAcc}
\end{figure}

\textbf{Utility analysis of private Algorithms.} Figure~\ref{fig:accuracy} shows the densities of private subgraphs in relative to the densities of the baselines.  In most networks which have average degree larger than $4$, we observe that \seqalgo{} has relative density $\geq 75\%$ for $\epsilon = 2 \text{ or } 4$. For $\epsilon = 2$,  \seqalgo{} has the same density as the non-private baseline in four of six networks and has at least $50\%$ of the density of the baseline in all except one. 
Figure~\ref{fig:densityVsAcc} shows the relationship between the relative density of the solution computed by our private mechanism and
the network density (i.e., the average degree) for the twenty networks we study. It confirms that, in general, the accuracy is higher in higher density graphs.
In general, we observe the trade-off between accuracy and privacy in all measurements. Higher $\epsilon$, which means less privacy guarantee, yields better accuracy. In contrast, Figures~\ref{fig:accuracy},~\ref{fig:jaccard},~\ref{fig:recall} show that $\delta$ does not have significant impacts on the quality of solutions. We show in the Appendix that,  in general, \palalgo~outperforms \phasealgo~at $\epsilon \geq 2$. For small $\epsilon$, neither algorithm has good accuracy, though \phasealgo~is slightly better.

\begin{figure}[h]
  \includegraphics[width=\columnwidth,keepaspectratio]{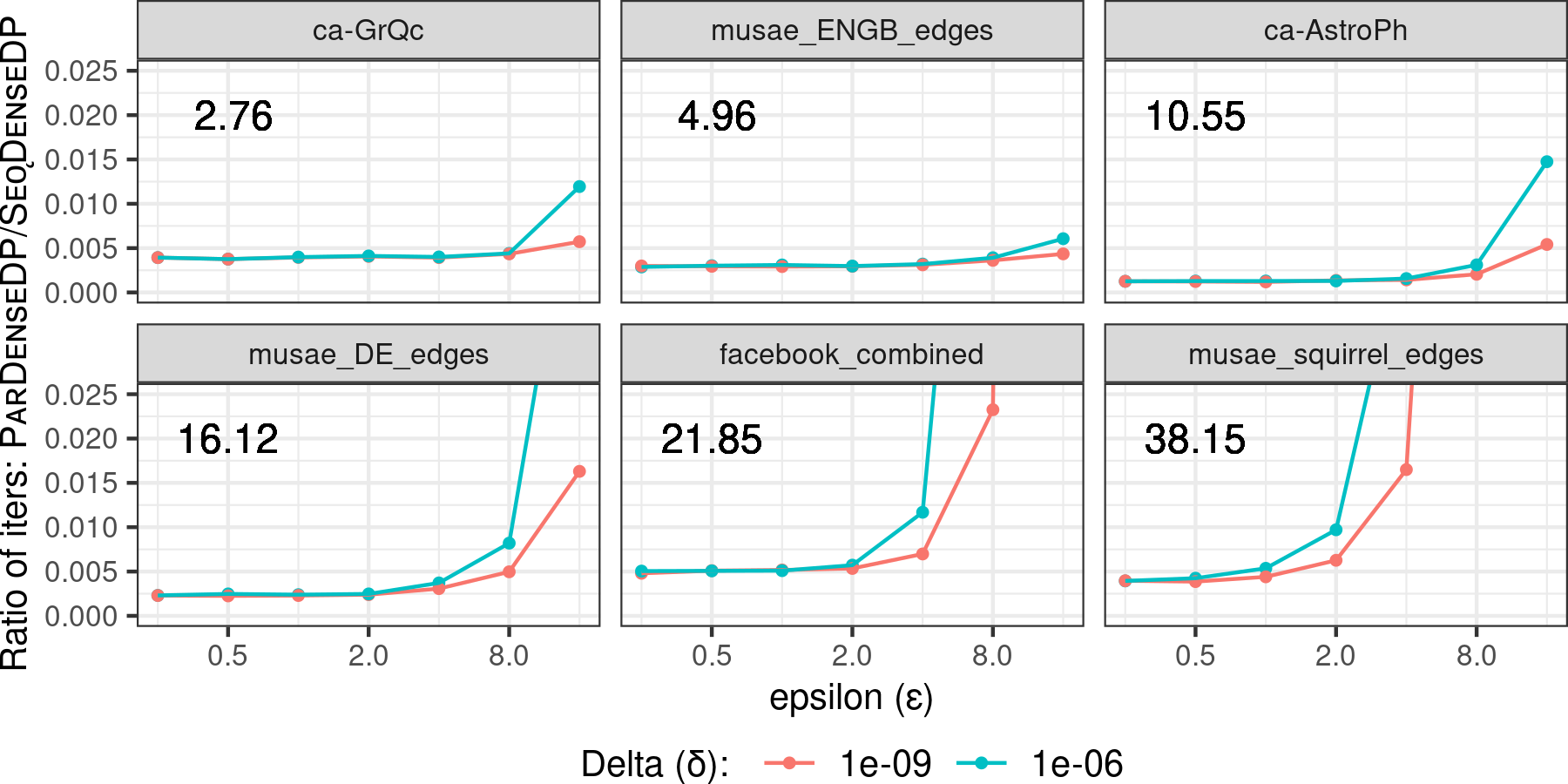}
  \centering
  \caption{\textbf{Number of iterations taken by \palalgo{}, relative to \seqalgo{}.} }
  \label{fig:parIters}
\end{figure}

\noindent
\textbf{Jaccard index and recall.}
Figure~\ref{fig:jaccard} shows the Jaccard similarity between the private subgraphs and $S_b$. The coefficients vary greatly across networks. Four out of the six networks have Jaccard similarity coefficients at least $0.5$ for $\epsilon \geq 2$ with \seqalgo. As with the relative density, networks with higher density tend to have better Jaccard similarity coefficients.
Figure~\ref{fig:recall} shows the recall of our algorithms, i.e., the fraction of nodes in $S_b$ which are selected by our private algorithms. The recall for \seqalgo{} is at least $75\%$ in all networks for $\epsilon \geq 1$, which indicates that the algorithm successfully finds most nodes in $S_b$.

\noindent
\textbf{Performance of parallel algorithms.} Figure~\ref{fig:parIters} shows the ratio of the number of iterations taken by \palalgo~and \seqalgo (recall that the number of iterations for \seqalgo{} is $n$). Due to the nature of the sampling probabilities in \palalgo{}, the probability of removing a node reduces as $\epsilon$ increases.
Consequently, \palalgo~takes more iterations to remove all nodes with larger $\epsilon$ and smaller $\delta$. 
In all but the \textit{musae\_squirrel}  network (Figure~\ref{fig:parIters}), the number of iterations for \palalgo{} is about $1\%$ of that for \seqalgo, when $\epsilon \leq 8$. Hence in most cases, \palalgo~is much more efficient.


\section{Conclusions}
\label{sec:conclusions}


In this paper, we design the first sequential and parallel differentially private algorithms for the densest subgraph problem in the edge-privacy model. All of them give $(1/2, O(\log{n}))$-approximate solutions, with high probability. In other words, they match the $1/2$ approximation of~\cite{charikar:dense}, 
with an additive approximation of $O(\log{n})$; we also prove a lower bound on the additive term of $\Omega(\sqrt{\log{n}})$. Our main technical contributions include adaptation of the exponential mechanism to be applied in parallel, and the analysis of privacy and accuracy. Our experiments on 20 networks show that our algorithms have good accuracy in high-density networks and reasonable recall, overall, when $\epsilon\geq 2$.  \seqalgo{} has better accuracy while the parallel variants reduce the number of iterations significantly in most settings. Our experiments suggest that \palalgo{} is good enough for practical uses among two parallel variants. Our paper leads to many open questions. 
Can we improve the runtime, e.g., by a combination of \palalgo{} and \phasealgo{}?
The most intriguing problem is to close the gap between the upper and lower bounds in accuracy. In particular, is an $O(\sqrt{\log{n}})$ additive approximation to~\cite{charikar:dense} possible to achieve?
Is it possible to obtain a $(1, O(\log{n}))$-approximate solution, i.e., a
purely additive approximation?

\vspace{3mm}
{\bf Acknowledgement. } We thank anonymous ICML reviewers for helpful suggestions.  A. Vullikanti is supported by NSF grants IIS-1931628, CCF-1918656,  NSF IIS-1955797, NIH grant R01GM109718.

\clearpage
\bibliography{ref}
\bibliographystyle{icml2021}

\onecolumn
\clearpage
\appendix
\section{Missing Proofs}

\subsection{\seqalgo{}}
\begin{lemma}
  \label{lemma:sequential-exponential-full}
The computation of the sequence $\pi^G = \pi_1,\ldots,\pi_n$, in the \textbf{for} loop in Algorithm~\ref{alg:private-densest} (lines 3--5) is $(\epsilon/2, \delta)$-differentially private.
\end{lemma}

\begin{proof}


Let $G\sim G'$ be two graphs that differ in exactly one edge $e=(u, v)$. Let $\pi^G$ and $\pi^{G'}$ denote the permutations computed in the for loop in Algorithm~\ref{alg:private-densest}. Let $\mathcal{T}$ is the first iteration that one of the endpoints of $e$ (i.e., $u$ or $v$) is removed. Let $adj(e)$ define the set of nodes adjacent to edge $e$.
Recall that $\degree{G}{j}$ is the degree of node $j$ in subgraph $S_t$ of graph $G$.

Fix a permutation $\pi$. Consider the following ratio:

\begin{eqnarray*}
  P &=& \frac{\Pr[\pi^G = \pi]}{\Pr[\pi^{G'} = \pi]} \\
    &=& \prod_{t=1}^{n} \frac{\exp(-\epsilon'\degree{G}{\pi_t})/\sum_{j}\exp(-\epsilon'\degree{G}{j})}{\exp(-\epsilon'\degree{G'}{\pi_t})/\sum_{j}\exp(-\epsilon'\degree{G'}{j})} \\
    &=& \frac{\exp(-\epsilon'\text{deg}_{S_\T}^G(\pi_\T))}{\exp(-\epsilon'\text{deg}_{S_\T}^{G'}(\pi_\T))} \prod_{t=1}^\T\frac{\sum_j \exp(-\epsilon'\degree{G'}{j})}{\sum_j \exp(-\epsilon'\degree{G}{j})} \\
\end{eqnarray*}

The last equality follows from the definition of $\T$, because: (i) for $t=\T + 1 ..n$, $\degree{G}{j}=\degree{G'}{j}$ for all nodes $j$, and, (ii) for $t = 1..\T-1$, $\degree{G}{\pi_t} = \degree{G'}{\pi_t}$. We have two cases for the analysis.

First, suppose, $G + e = G'$. In this case, we have $\frac{\exp(-\epsilon'\text{deg}_{S_\T}^G(\pi_\T))}{\exp(-\epsilon'\text{deg}_{S_\T}^{G'}(\pi_\T))} = \exp(\epsilon')$ because $\text{deg}_{S_\T}^G(\pi_\T) + 1 = \text{deg}_{S_\T}^{G'}(\pi_\T)$. Further, for each $t, j$, we have  $\degree{G'}{j} \geq \degree{G}{j}$ for all $j$, which implies the product $\prod_{t=1}^\T\frac{\sum_j \exp(-\epsilon'\degree{G'}{j})}{\sum_j \exp(-\epsilon'\degree{G}{j})}\leq 1$. Therefore, $P \leq \exp(\epsilon')$

In the second case, $G - e = G'$. Similar to the first case, $\frac{\exp(-\epsilon'\text{deg}_{S_\T}^G(\pi_\T))}{\exp(-\epsilon'\text{deg}_{S_\T}^{G'}(\pi_\T))} = \exp(-\epsilon') \leq 1$. 
For $\jAdjE, t\leq \T$, we have $\text{deg}_{S_t}^{G}(j) = \text{deg}_{S_t}^{G'}(j)+1$. For $\jnotAdjE$, and for all $t$, $\degree{G}{j} = \degree{G'}{j}$. Further, for $t>\T$, we have $\degree{G}{j} = \degree{G'}{j}$ for all $j$. Note that whenever we use $\degree{G}{j}$, we only consider $j\in S_{t-1}$, i.e., $j$ would not have been deleted from the graph before step $t$.


Therefore, for $t\leq\T$:
\begin{eqnarray*}
  &&\sum_j\exp(-\epsilon'\degree{G'}{j}) \\
  &=&  \sum_\jAdjE\exp(-\epsilon'\degree{G'}{j}) + \sum_\jnotAdjE\exp(-\epsilon'\degree{G'}{j})\\
                                      &=& \sum_\jAdjE\exp(-\epsilon'(\degree{G}{j} - 1)) + \sum_\jnotAdjE\exp(-\epsilon'\degree{G}{j})\\
                                      &=& \exp(\epsilon')\sum_\jAdjE\exp(-\epsilon'\degree{G}{j}) + \sum_\jnotAdjE\exp(-\epsilon'\degree{G}{j})\\
                                      &=& (\exp(\epsilon') - 1)\sum_\jAdjE\exp(-\epsilon'\degree{G}{j}) + (\sum_\jAdjE\exp(-\epsilon'\degree{G}{j}) + \sum_\jnotAdjE\exp(-\epsilon'\degree{G}{j}))\\
                                      &=& (\exp(\epsilon') - 1)\sum_\jAdjE\exp(-\epsilon'\degree{G}{j}) + \sum_j \exp(-\epsilon'\degree{G}{j})\\
\end{eqnarray*}

We now return to the expression for $P$. Since $\degree{G}{j} = \degree{G'}{j}$ for all $j$ and $t>\T$, we have
\begin{eqnarray*}
  P &\leq& \prod_{t = 1}^\T (\frac{(\exp(\epsilon') - 1)\sum_\jAdjE\exp(-\epsilon'\degree{G}{j}) }{\sum_j \exp(-\epsilon'\degree{G}{j})}\\
  && + \frac{\sum_j \exp(-\epsilon'\degree{G}{j})}{\sum_j \exp(-\epsilon'\degree{G}{j})}) \\
    &=& \prod_{t = 1}^T (1 + (\exp(\epsilon') - 1)\frac{\sum_\jAdjE\exp(-\epsilon'\degree{G}{j})}{\sum_j\exp(-\epsilon'\degree{G}{j})}) \\
        &=&\prod_{t = 1}^T (1 + (\exp(\epsilon') - 1)p_t(G))\\
    &\leq& \prod_{t=1}^{\T}\exp((\exp(\epsilon')-1)p_t(G)),\\
           &&\mbox{ using the inequality $1+x\leq e^x$ for all $x\geq 0$},
\end{eqnarray*}

where $p_t(G)$ is the probability that an endpoint of edge $e$ is picked in iteration $t$. We will show in Lemma~\ref{lemma:good}, by using the adversarial process (Section~\ref{sec:dp-basic}), that 
$\sum_{t=1}^{\T-1}p_t(G) \leq \ln\delta^{-1}$ with probability at least $1 - \delta$. Assuming this bound, we first derive a bound on $P$. We say that $\pi$ is $\ln{\delta^{-1}}$-good if $\sum_{t=1}^{\T-1}p_t(G) \leq \ln\delta^{-1}$ and $\ln{\delta^{-1}}$-bad otherwise. Therefore, with probability at least $1-\delta$, the ordering $\pi$ is $\ln{\delta^{-1}}$-good and

\begin{eqnarray*}
  P &\leq& \prod_{t=1}^{\T}\exp((\exp(\epsilon')-1)p_t(G)) \\
    &\leq& \exp(2\epsilon'\sum_{t=1}^\T p_t(G)), \\ && \mbox{ using the inequality $e^x\leq 1+2x$ for $x\leq 1$} \\
    &\leq& \exp(2\epsilon'(\ln\delta^{-1} + p_\T(G))) \\
    &\leq& \exp(2\epsilon'(\ln\delta^{-1}+1)) \\
    &=& \exp(2\epsilon'(\ln(e/\delta)) \\
    &=& \exp(\epsilon/2), \mbox{ since $\epsilon' = \frac{\epsilon}{4\ln{(e/\delta)}}$},
\end{eqnarray*}
which implies $\Pr[\pi^G = \pi] \leq exp(\epsilon/2)\Pr[\pi^{G'}=\pi]$ in this case.

Let $\mathcal{P}$ be the set of output orderings that might be generated by the \textbf{for} loop in \seqalgo. Then, we have
\begin{eqnarray*}
 && \Pr[\pi^{G} \in \mathcal{P}] \\ &=& \sum_{\pi\in\mathcal{P}}\Pr[\pi^{G} = \pi]\\
    &=& \sum_{\pi\in\mathcal{P}:\mbox{$\pi$ is $(\ln{\delta^{-1}})$-good}}\Pr[\pi^{G} = \pi]  
    + \sum_{\pi\in\mathcal{P}:\mbox{$\pi$ is $(\ln{\delta^{-1}})$-bad}}\Pr[\pi^{G} = \pi] \\             &\leq& \sum_{\pi\in\mathcal{P}:\mbox{$\pi$ is $(\ln{\delta^{-1}})$-good}}\exp(\epsilon/2)\Pr[\pi^{G'} = \pi]
    + \delta\\
  &\leq&\exp(\epsilon/2)\Pr[\pi^{G'}\in\mathcal{P}] + \delta
\end{eqnarray*}
The lemma then follows.
\end{proof}

\subsection{\palalgo{}}
\begin{lemma}
  \label{lemma:Etj}
Let $G=G'+e$, where $e=(u, v)$. Let $E_{tj}= \frac{1-P_{tj}^G}{1-P_{tj}^{G'}}$ be the ratio of the probability that node $j$ is not picked in iteration $t$ in $G$, to the corresponding probability in $G'$. Let $p^{uv}_t(G)$ be the probability that at least one of the nodes $u$ or $v$ is removed in iteration $t$ from graph $G$.
Then, for $\jAdjE$, $E_{tj}\leq\exp(\frac{e^{\epsilon'}-1}{1-e^{-1}}P_{tj}^G)$, and $E_{tu}E_{tv}\leq \exp(\frac{4\epsilon'}{1-e^{-1}}p^{uv}_t(G))$.
\end{lemma}

\begin{proof}
For $\jAdjE$, we have $\text{deg}^G_{S_t}(j) = \text{deg}^{G'}_{S_t}(j)+1$. Therefore,

\begin{eqnarray*}
E_{tj} &=& \frac{1-P^G_{tj}}{1-\exp(\epsilon')P^{G}_{tj}}\\
&=& \frac{1-P^G_{tj}(\exp(\epsilon') + 1 - \exp(\epsilon'))}{1-\exp(\epsilon')P^{G}_{tj}}\\
&=& \frac{1-P^G_{tj}\exp(\epsilon') - (1 - \exp(\epsilon'))P^G_{tj}}{1-\exp(\epsilon')P^{G}_{tj}}\\
&=& 1 + \frac{(\exp(\epsilon') - 1)P^G_{tj}}{1-\exp(\epsilon')P^{G}_{tj}}\\
&\leq& 1 + \frac{(\exp(\epsilon') - 1)P^G_{tj}}{1-e^{-1}}\\
&\leq& \exp\left(\frac{\exp(\epsilon') - 1}{1-e^{-1}}P^G_{tj}\right),
\end{eqnarray*}
where the second last inequality follows from the following argument
\begin{eqnarray*}
deg_{S_t}(j) + c &\geq& \frac{1}{\epsilon'} + 1\\
\Rightarrow \epsilon'(deg_{S_t}(j) + c) - \epsilon' &\geq& 1\\
-\epsilon'(deg_{S_t}(j) + c) + \epsilon' &\leq& -1\\
P^G_{tj} e^{\epsilon'} = exp(-\epsilon'(deg_{S_t}(j) + c)) e^{\epsilon'} &\leq& e^{-1}\\
\Rightarrow 1 - P^G_{tj} e^{\epsilon'} &\geq& 1 - e^{-1}\\
\Rightarrow \frac{1}{1 - P^G_{tj} e^{\epsilon'}} &\leq& \frac{1}{1 - e^{-1}},
\end{eqnarray*}


where the last inequality requires $P^G_{tj}e^{\epsilon'}-1 < 1-e^{-1}$. This holds because $P^G_{tj} \leq \exp(-\epsilon'-1) \Rightarrow P^G_{tj}e^{\epsilon'} -1 \leq e^{-1} -1 < 1 - e^{-1}$.
This implies
\begin{eqnarray*}
E_{tu} E_{tv} &\leq& \exp\left(\frac{\exp(\epsilon')-1}{1-e^{-1}}(P^G_{tu} + P^G_{tv})\right)\\
              &\leq& \exp\left(\frac{\exp(\epsilon')-1}{1-e^{-1}}2p_t(G)\right)\\
              &\leq& \exp\left(\frac{4\epsilon'}{1-e^{-1}}p_t(G)\right),
\end{eqnarray*}
where the second inequality follows because $\max(P^G_{tu}, P^G_{tv}) \leq p_t(G)$, where $p_t(G)$ is the probability that at least one of $u$ or $v$ is removed at step $t$, i.e., $p_t(G) = P_{tu} + P_{tv} - P_{tu}P_{tv}$. The last inequality follows because $e^{\epsilon'} < 1+2\epsilon'$, if $\epsilon'\leq 1$. 
\end{proof}

\begin{lemma}
  \label{lemma:parallel-exponential}
  The computation of the sequence $S_1, S_2, \ldots$ by the \textbf{while} loop (line 4-8) of \palalgo{} is $(\epsilon/2, \delta)$-differentially private.
\end{lemma}

\begin{proof}
Let $G \sim G'$ be two neighboring graphs which differ in a single edge $e=(u, v)$. Let $\pi^G$ and $\pi^{G'}$ denote the outputs in the case of graphs $G$ and $G'$, respectively. As in the proof of Lemma~\ref{lemma:sequential-exponential}, let $\mathcal{T}$ be the first iteration in which a node in ${u, v}$ is removed, and let $P_{tj}^G$ be the probability that node $j$ of graph $G$ is sampled at step $t$. We note that for every node $j\neq u, v$, in every iteration $t$: $deg_{S_t}^G(j) = deg_{S_t}^{G'}(j)$ and hence $P_{tj}^G = P_{tj}^{G'}$. Fix a permutation $\pi$, and consider the following ratio:

\begin{eqnarray*}
  P &=& \frac{\Pr[\pi^G = \pi]}{\Pr[\pi^{G'} = \pi]} \\
    &=& \prod_{t=1}^\T\frac{\Pr[\pi_t^G = \pi]}{\Pr[\pi_t^{G'} = \pi]}, \substack{\text{ since }\degree{G}{j} = \degree{G'}{j} \\ \text{ for }t> \T, \text{for all }j} \\
    &=& \prod_{t=1}^\T\frac{\prod_{j: p_{tj}=1}P_{tj}^G \times \prod_{j: p_{tj}=0}(1 - P_{tj}^G)}{\prod_{j: p_{tj}=1}P_{tj}^{G'} \times \prod_{j: p_{tj}=0}(1 - P_{tj}^{G'})} \\
    &=&\prod_{t=1}^{\T-1}(\prod_{j: p_{tj}=1}\frac{P_{tj}^G}{P_{tj}^{G'}} \times \prod_{j: p_{tj}=0}\frac{1-P_{tj}^G}{1-P_{tj}^{G'}}) 
    \times \prod_{j: p_{\T j}=1}\frac{P_{\T j}^G}{P_{\T j}^{G'}} \times \prod_{j: p_{\T j}=0}\frac{1-P_{\T j}^G}{1-P_{\T j}^{G'}} \\
\end{eqnarray*}
\[
\mbox{Define } A_t = \prod_{j: p_{tj}=1}\frac{P_{tj}^G}{P_{tj}^{G'}}, 
B_t = \prod_{j: p_{tj}=0}\frac{1-P_{tj}^G}{1-P_{tj}^{G'}},\]\[
  C = \prod_{j: p_{\T j}=1}\frac{P_{\T j}^G}{P_{\T j}^{G'}},
  D = \prod_{j: p_{\T j}=0}\frac{1-P_{\T j}^G}{1-P_{\T j}^{G'}}\\
\]

We have two cases: $G + e = G'$ and $G - e = G'$.

\noindent
\textbf{Case 1} $G + e= G'$. For all $t$, $A_t=1$ because for all nodes $j$ and $t<\T$, $\degree{G}{j} = \degree{G'}{j}$. Next, 
\[
B_t = \prod_{j: p_{tj}=0}\frac{1-P_{tj}^G}{1-P_{tj}^{G'}} = \frac{(1-P_{tu}^G)(1-P_{tv}^{G})}{(1-P_{tu}^{G'})(1-P_{tv}^{G'})} \leq 1,
\]
because $P_{tj}^{G'} = \exp(-\epsilon')P_{tj}^{G} \leq P_{tj}^{G}$ for $j\in \{u, v\}$ and $P_{tj}^{G'} = P_{tj}^{G}$ for $j\notin \{u, v\}$. Therefore, $\prod_{t=1}^{\T -1} (A_t\times B_t) \leq 1$. We have two sub-cases.

\noindent
\textbf{Case 1.a} Both $u, v$ are picked at step $\T$: $p_{\T u} = p_{\T v} = 1$. In that case, none of the terms in $D$ involve $u$ or $v$, which means $D = 1$. We have

\begin{align*}
C &= \prod_{j: p_{\T j}=1}\frac{P_{\T j}^G}{P_{\T j}^{G'}}
  = \frac{P_{\T u}^G\times P_{\T v}^{G}}{P_{\T u}^{G'}\times P_{\T v}^{G'}} \\
  &= \frac{P_{\T u}^G\times P_{\T v}^{G}}{\exp(-2\epsilon')\times P_{\T u}^{G}\times P_{\T v}^{G}} 
= \exp(2\epsilon'),
\end{align*}
since $\text{deg}^{G'}_{S_\T}(j) =\text{deg}^G_{S_\T}(v)+1$, for $\jAdjE$. Therefore, $P = \prod_{t = 1}^{\T-1}(A_t\times B_t)\times C \times D \leq \exp(2\epsilon')$.

\noindent
\textbf{Case 1.b} Exact one node in $\{u, v\}$ is picked at step $\T$. Assume $v$ is picked, so that $p_{\T u} = 0, p_{\T v} = 1$, and $C = exp(\epsilon')$ and $D = \frac{1-P_{\T u}^G}{1-P_{\T u}^{G'}} \leq 1$. Therefore, $P \leq \exp(\epsilon')$ in this case as well, which implies that in \textbf{Case 1}, $P \leq \exp(2\epsilon') < \exp(\epsilon/2)$. 

\noindent
\textbf{Case 2.} $G = G' + e$: by the same argument as in Case 1, we have $A_t = 1$ for all $t<\T$. We have similar two sub-cases.

\noindent 
\textbf{Case 2.a} Both $u, v$ are picked at step $\T$: then $p_{\T u} = p_{\T v} = 1$, and $C = exp(-2\epsilon') \leq 1$ and $D = 1$. This implies $P \leq \prod_{t =1}^{\T -1}B_t$

\noindent 
\textbf{Case 2.b} Only one of $u, v$ is picked at step $\T$. Assume $v$ is picked, so that $p_{\T u} = 0, p_{\T v} = 1$. Then $C = exp(-\epsilon') \leq 1$, $D = \frac{1-P_{\T u}^G}{1-P_{\T u}^{G'}}$, and $P \leq \prod_{t =1}^{\T -1}B_t \times  \frac{1-P_{\T u}^G}{1-P_{\T u}^{G'}}$

Now, we consider the product $B_t$. Since for all $j\not\in \text{adj}(e)$, $\degree{G}{j} = \degree{G'}{j}$, the terms corresponding to such nodes cancel out in $B_t$, and we have
\begin{eqnarray*}
  \prod_{t=1}^{\T-1}B_t &=& \prod_{t=1}^{\T-1}\frac{(1-P_{tu}^G)(1-P_{tv}^{G})}{(1-P_{tu}^{G'})(1-P_{tv}^{G'})} \\
                        &=& \prod_{t=1}^{\T-1}(E_{tu} \times E_{tv}) \\      
    &\leq& \prod_{t=1}^{\T-1}\exp\left(\frac{4\epsilon'}{1-e^{-1}}p^{uv}_t(G)\right) \mbox{ from Lemma~\ref{lemma:Etj}}\\
     &=& \exp\left(\frac{4\epsilon'}{1-e^{-1}}\sum_{t=1}^{\T-1}p^{uv}_t(G)\right). \\
    \end{eqnarray*}

As in the proof of Lemma~\ref{lemma:sequential-exponential}, we will prove later that
$\sum_{t=1}^{\T -1 } p_t(G)\leq \ln(1/\delta)$. with probability at least $1-\delta$. Assuming this bound, we complete the rest of proof. With probability at least $1-\delta$, we have $\prod_{t=1}^{\T-1}B_t \leq \exp\left(\frac{4\epsilon'}{1-e^{-1}}\ln{(1/\delta)}\right)$. We now apply this inequality to both cases 2.a and 2.b.

\noindent
\textbf{Case 2.a} $P = \prod_{t=1}^{\T-1}B_t \leq \exp\left(\frac{4\epsilon'}{1-1/e}\ln(1/\delta)\right)\leq\exp(\epsilon/2)$.

\noindent
\textbf{Case 2.b} $P = \prod_{t=1}^{\T-1}B_t \times D$
\begin{eqnarray*}
  P &=& \prod_{t=1}^{\T-1}B_t \times \frac{1-P_{\T u}^G}{1-P_{\T u}^{G'}} \\
    &\leq&\exp\left(\frac{4\epsilon'}{1-1/e}\ln(1/\delta)\right) \times E_{\T u} \\
    &\leq&\exp\left(\frac{4\epsilon'}{1-1/e}\ln(1/\delta)\right) \times \exp\left(\frac{4\epsilon'}{1-1/e}p_{\mathcal{T}}(G)\right) \\
    &\leq&\exp\left(\frac{4\epsilon'}{1-1/e}\ln(1/\delta)\right) \times \exp\left(\frac{4\epsilon'}{1-1/e}\right) \\
    &=&\exp\left(\frac{4\epsilon'}{1-1/e}(\ln(1/\delta)+1)\right) \\
    &=&\exp\left(\frac{4\epsilon'}{1-1/e}\ln(e/\delta)\right) \\
    &=&\exp(\epsilon/2), \\
\end{eqnarray*}

where the second inequality follows because of Lemma~\ref{lemma:Etj}.

Therefore, in \textbf{Case 2}, we have $P \leq \exp(\epsilon/2)$, with probability at least $1-\delta$. Using the same argument as in the proof of Lemma~\ref{lemma:sequential-exponential}, we have that for any set of outputs $\mathcal{P}$ that can be generated by the \textbf{while} loop of \palalgo{}, $\Pr[\pi^G\in\mathcal{P}] \leq \exp(\epsilon/2)\Pr[\pi^{G'}\in\mathcal{P}]+\delta$, and hence the lemma follows.

Finally, we prove the upper bound of the sum $\sum_{t=1}^{\T-1}p^{uv}_t(G) \text{ is } \ln(1/\delta)$, with probability at least $1-\delta$, by mapping our algorithm to the adversarial probabilistic process in Lemma~\ref{lemma:adversarial}. The adversarial process is defined as following: in each iteration $t$, the adversary chooses heads with probability equal to $p^{uv}_t(G)$. If a head is tossed, each node $j\in\{u, v\}$ is picked independently with probability $P_{tj}$. If a tail is tossed, the adversary does not pick any node of $\{u, v\}$. Every other node $j\not\in\{u, v\}$ is picked independently with probability $P_{tj}$, no matter how the coin is tossed. Therefore, in case of heads, the outcomes are $\{(u), (v), (u, v)\}$ (which means picking only $u$, picking only $v$ and picking both $u$ and $v$), with probabilities proportional to $P_{tu}-P_{tu}P_{tv}$, $P_{tv}-P_{tu}P_{tv}$ and $P_{tu}P_{tv}$, respectively. In order to prove the adversarial process and our algorithm are equivalent, we have to prove that probabilities of picking a node $j$ are the same in both processes.

For $j\notin\{u, v\}$, the adversarial picks $j$ with probability $P_{tj}$, so the probability of picking $j$ is the same as in Algorithm~\ref{algo:parallel}.

We consider the probability of picking $u$ ($v$ is similar) at iteration $t$ is $\Pr[u \text{ is picked}]$ :
\begin{eqnarray*}
   &=& \Pr[\text{head}](\Pr[(u)|\text{head}] + \Pr[(u,v)|\text{head}])\\
  &=& p_t(G)\frac{(P_{tu}-P_{tu}P_{tv})+(P_{tu}P_{tv})}{P_{tu}-P_{tu}P_{tv}+ P_{tv}-P_{tu}P_{tv} + P_{tu}P_{tv}} \\
   &=& (P_{tu} + P_{tv} - P_{tu}P_{tv})\frac{P_{tu}}{P_{tu}+ P_{tv}-P_{tu}P_{tv}} \\
  &=& P_{tu},.
\end{eqnarray*}
which is the same with probability of picking $u$ in Algorithm~\ref{algo:parallel}

Similarly, we also have the probability that $v$ picked at iteration $t$ is $P_{tv}$. Therefore, the two processes (our algorithm and the adversarial probabilistic process) are equivalent. 
By Corollary~\ref{cor:adversarial} that $\sum_{t=1}^{\T -1 } p_t(G)\leq \ln(1/\delta)$ with probability at least $1-\delta$, we have: 
\[
  \prod_{t=1}^{\T-1}B_t \leq \exp\left(\frac{4\epsilon'}{1-e^{-1}}\ln{(1/\delta)}\right), 
\]

\end{proof}

\begin{theorem}
  \label{theorem:parallel-alg-full}
  Algorithm~\ref{algo:parallel} is $(\epsilon, \delta)$-differentially private.
\end{theorem}
 \begin{proof}
By Lemma~\ref{lemma:parallel-exponential}, the sequence $\{S_1, S_2, ...\}$ output by the \textbf{while} loop is $\left(\epsilon/2, \delta\right)$-differentially private. For each $S_t$, adding or removing 1 edge to the graph changes $\rho(S_t)$ by at most 1, hence $\Delta_{\rho} = 1$. Applying the Exponential Mechanism, the last step of \palalgo{} releases output $S$ with $\epsilon/2$-differential privacy. Using the Composition Theorem, Algorithm~\ref{algo:parallel} preserves $(\epsilon, \delta)$-differential privacy.
 \end{proof}

\begin{theorem}
\label{theorem:parallel-accuracy-full}
(Theorem~\ref{theorem:parallel-accuracy})
Let $S^*$ be an optimal solution. If $\delta\leq 1/e$,
the set $S$ output by Algorithm~\ref{algo:parallel} satisfies $\rho(S)\geq \frac{\rho(S^*)}{2} - \frac{56}{\epsilon}\ln(1/\delta)\ln{n}$, with probability at least $1-2/n$.  
\end{theorem}
 \begin{proof}
 In each iteration $t$, observe that if a node $j$ has $deg_{S_t}(j)\geq 2\rho(S_t)+\frac{3}{\epsilon'}\ln{n}\geq \frac{3}{\epsilon'}\ln{n}$, then $\Pr[p_{tj}=1]\leq exp(-\epsilon'(deg_{S_t}(j) + c))\leq exp(-\epsilon'deg_{S_t}(j))\leq 1/n^3$. Therefore, with probability at least $1-1/n$, for all $t$ and $j$, $deg_{S_t}(j)\leq 2\rho(S_t)+\frac{3}{\epsilon'}\ln{n}$. Suppose $T$ is the first iteration in which a node $v\in S^*\cap S_T$ is picked. Then, As in the proof of Theorem~\ref{theorem:approx-seq}, we have $\rho(S^*)\leq 2\rho(S_T) + \frac{3}{\epsilon'}\ln{n}$, which implies $\rho(S_T)\geq \rho(S^*)/2 - \frac{3}{2\epsilon'}\ln{n}$.
 
 Next, in the last line of the algorithm, suppose $S_{\tau}=\mbox{argmax}_{S_t} \rho(S_t)$. Selecting the distinct set of $S_t$ is a post-processing of the private list of all $S_t$ and hence, the set is private and has at most $n$ elements. As in the proof of Theorem~\ref{theorem:approx-seq}, we have $\rho(S)\geq \rho(S_{\tau}) - \frac{8}{\epsilon}\ln{n}\geq \rho(S^*)/2 - \frac{12\ln(e/\delta)}{(1-1/e)\epsilon}\ln{n} - \frac{8}{\epsilon}\ln{n} \geq \rho(S^*)/2 - \left(\frac{24}{1-1/e}+\frac{8}{\ln(1/\delta)}\right)\frac{1}{\epsilon}\ln(1/\delta)\ln(n) \geq
 \rho(S^*)/2 - \left(\frac{24}{1-1/2}+8\right)\frac{1}{\epsilon}\ln(1/\delta)\ln(n) \geq
 \rho(S^*)/2 - \frac{56}{\epsilon}\ln(1/\delta)\ln(n)
 $, with probability at least $1-2/n$.
 \end{proof}

\subsection{\phasealgo{}}
\begin{lemma}
  \label{lemma:phase-density}
For each phase $i$, except the last, with probability at least $1-1/n^4$, $\rho(S_i)\leq \widehat{\rho(S_i)}\leq \rho(S_i) + \frac{32}{\epsilon}\ln{n}$  
\end{lemma}
\begin{proof}
By the property of the Laplace distribution~\cite{dwork:fttcs14}, $\Pr\left[|Lap(\frac{4\ln{n}}{|S_i|\epsilon})|>\frac{16\ln^2{n}}{|S_i|\epsilon}\right]\leq exp(-4\ln{n}) = 1/n^4$. Therefore, with probability at least $1-1/n^4$, we have $|Lap(\frac{4\ln{n}}{|S_i|\epsilon})|\leq \frac{16\ln^2{n}}{|S_i|\epsilon} \leq \frac{16\ln{n}}{\epsilon}$, since $|S_i|>\ln{n}$ for all $i$, except the last phase. This implies that with probability at least $1-1/n^4$, we have $-\frac{16\ln{n}}{\epsilon}\leq Lap(\frac{4\ln{n}}{|S_i|\epsilon}) \leq \frac{16\ln{n}}{\epsilon}$. By the definition of $\widehat{\rho(S_i)}$, the Lemma follows.
\end{proof}

\begin{lemma}
\label{lemma:parallel-node-maxdeg}
With probability at least $1-\frac{4\ln{n}}{n^2}$, in every phase $i$ (except the last one), for any node $v$ with $d_{S_i}(v) > 4\widehat{\rho(S_i)} + \frac{4}{\epsilon'}\ln{n}$, we have $T_v > T_i$.
\end{lemma}
\begin{proof}
Consider any phase $i$, and node $v\in S_i$. We have $P_v = exp(-\epsilon'(d_{S_i}(v)+c)) \leq exp(-\epsilon'(4\widehat{\rho(S_i)}+c) - 4\ln{n}) = exp(-\epsilon'(4\widehat{\rho(S_i)}+c))/n^4$, in which the inequality is because $d_{S_i}(v) > 4\widehat{\rho(S_i)} + \frac{4}{\epsilon'}\ln{n}$. Since $T_v$ is sampled from a geometric process, we have $\Pr[T_v\leq T_i] = \sum_{j=1}^{T_i} (1-P_v)^{j-1}P_v\leq T_i P_v \leq \frac{4\ln{n}}{n^4}$. By taking a union bound over all nodes, and all phases, the lemma follows.
\end{proof}

\begin{lemma}
\label{lemma:process-phases}
In Algorithm~\ref{algo:parallel-phases}, let $p_t(G)$ denote the probability that one of $u, v$ is removed at time $t$. Let $q^u_t(G)$ denote the probability that node $u$ is removed at time $t$, similarly $q^v_t(G)$. Let $\T_u,\T_v$ are iterations in which $u$ and $v$ are removed respectively. Assume $\T_u < \T_v$, with probability at least $1-\delta/2$, $\sum_{t=1}^{\T_u-1} p_t(G)\leq \ln{4/\delta}$ and $\sum_{t=\T_u}^{\T_v-1} q^v_t(G)\leq \ln{4/\delta}$.
\end{lemma}

\begin{proof}
We use the same notations as in Lemma~\ref{lemma:parallel-exponential}. Let $\T_u \leq \T_v$ are iterations when each node in $\{u, v\}$ is removed. Let $\delta' = \delta/4$.

For iteration $t \in [1, \T_u]$, set $p_t(G) = P_{tu} + P_{tv} - P_{tu}P_{tv}$ as the same as in the random process of Lemma~\ref{lemma:parallel-exponential}. Therefore, the algorithm from iteration $1 \rightarrow \T_1$ is equivalent to 1 instance of the Adversarial Probabilistic Process. Applying Corollary~\ref{cor:adversarial}, with probality at least $1-\delta'$, $\sum_{t=1}^{\T_u-1} p_t(G)\leq \ln{1/\delta'} = ln{4/\delta}$.

In the second process after $u$ is removed ($t \in [\T_u+1, \T_v]$), for all node $j \neq v$, the probabilities of picking $j$ are set to be independent from the coin tosses, hence their probabilities are the same. At each iteration $t \in [\T_u+1, \T_v]$, set $t' = t - \T_u$, the adversary chooses head probability $q^v_{t'}(G) = P_{tv}$ and picks $v$ if a head is tossed and does not pick $v$ if a tail is tossed. Then, the probability to pick $v$ at step $t$ is still $P_{tv}$ and the sampling process is equivalent to another instance of the Adversarial Probabilistic Process. Applying Corollary~\ref{cor:adversarial}, with probability at least $1-\delta'$, $\sum_{t=\T_u}^{\T_v-1} q^v_t(G) = \sum_{t'=1}^{\T_v-\T_u-1} q^v_{t'}(G)\leq \ln{1/\delta'}= \ln{4/\delta}$.

Substituting $\delta = 4\delta'$, with probability at least $(1-\delta')^2 \geq {1-\delta/2}$, $\sum_{t=1}^{\T_u-1} p_t(G)\leq \ln{4/\delta}$ and $\sum_{t=\T_u}^{\T_v-1} q^v_t(G)\leq \ln{4/\delta}$.
\end{proof}

\begin{lemma}
  \label{lemma:phase-privacy}
  The computation of the list $\{S_0, S_1,\ldots\}$ by the outer \textbf{while} loop is $(\epsilon/2, \delta)$-differentially private, with $\epsilon\in(0,1]\text{ and }\delta\in(2n^{-2},1)$.
\end{lemma}

\begin{proof}

We outline the proof as following: we prove that the sequence $\{\pi_0, \pi_1,\ldots\}$ is produced with $(\epsilon/4, \delta/2)$-differential privacy, no matter when we start a new phase. At each phase $i$, we only use private information derived from private $\widehat{\rho(S_i)}$ to end a phase and calculate $S_i$. Next, we prove that the sequence $\{\widehat{\rho(S_0)}, \widehat{\rho(S_1)}, \ldots\}$ is also $(\epsilon/4, \delta/2)$-differentially private. In the end, the sequence $\{S_0, S_1,\ldots\}$ is $(\epsilon/2, \delta)$-differentially private due to the Composition Theorem by composing $\{\pi_0, \pi_1,\ldots\}$ and $\{\widehat{\rho(S_0)}, \widehat{\rho(S_1)}, \ldots\}$.

First, we prove that $\{\pi_0, \pi_1,\ldots\}$ is produced with $(\epsilon/4, \delta/2)$-differential privacy.

We use the same notations and analyses of Lemma~\ref{lemma:parallel-exponential}. \textbf{Case 1.a} and \textbf{Case 2.a} remain unchanged, when both node $u$ and $v$ are removed at the same iteration $\T=\T_u=\T_v$. We have to re-evaluate \textbf{Case 1.b} and \textbf{Case 2.b}, when only one of $\{u, v\}$ is removed at iteration $\T$, and the other is removed later. We assume that $u$ is removed first at $\T=\T_u$ and $v$ is removed at $\T_v > \T_u$. The reason for the differences is in Lemma~\ref{lemma:parallel-exponential}, after $\T_u$, we update the degree of $v$ with respect to the remaining nodes and hence, this quantity is the same in both $G$ and $G'$, at every remaining iteration. In Algorithm~\ref{algo:parallel-phases}, we do not update the degree of $v$ until the end of the current phase. It means that in the worst case, degree of $v$ remains the same until it is removed, or $\forall{t\in\{\T_u, \T_v\}},  P_{tv}^G \leq \exp(\epsilon')P_{tv}^{G'}$ if $G' = G + \{u, v\}$. 

Recall that all terms are the same in $G$ and $G'$ after $v$ is removed:

\begin{eqnarray*}
P    &=&\prod_{t=1}^{\T_v-1}(\prod_{j: p_{tj}=1}\frac{P_{tj}^G}{P_{tj}^{G'}} \times \prod_{j: p_{tj}=0}\frac{1-P_{tj}^G}{1-P_{tj}^{G'}}) 
    \times \prod_{j: p_{\T_v j}=1}\frac{P_{\T_v j}^G}{P_{\T_v j}^{G'}} \times \prod_{j: p_{\T_v j}=0}\frac{1-P_{\T_v j}^G}{1-P_{\T_v j}^{G'}} \\
\end{eqnarray*}
\[\text{Let } A_t = \prod_{j: p_{tj}=1}\frac{P_{tj}^G}{P_{tj}^{G'}}, 
  B_t = \prod_{j: p_{tj}=0}\frac{1-P_{tj}^G}{1-P_{tj}^{G'}},
\]
\[
  C = \prod_{j: p_{\T_v j}=1}\frac{P_{\T_v j}^G}{P_{\T_v j}^{G'}}, 
  D = \prod_{j: p_{\T_v j}=0}\frac{1-P_{\T_v j}^G}{1-P_{\T_v j}^{G'}}
\]

\textbf{Case 1.b}. $G + (u, v) = G'$.

Product $A_t$ involves all terms of nodes $j$ picked in iteration $t$. All terms involve in this product are the same for all nodes except for $u$ (which is picked at $\T_u$), hence $A_t = 1$ for all $t\neq \T_u$, and $A_{\T_u}=\frac{P^{G}_{\T u}}{P^{G'}_{\T u}}=\exp(\epsilon')$.


Product $B_t$ contains terms of remaining nodes that are not picked at iteration $t$. Before $\T_u$, the numerator and denominator differ by the probabilities of picking $u$ and $v$ in $G$ and $G'$, while the other terms are the same. After iteration $\T_u$, because $u$ is already removed, there is only $v$'s term left in the difference. $B_t = \frac{(1-P_{tu}^G)(1-P_{tv}^{G})}{(1-P_{tu}^{G'})(1-P_{tv}^{G'})}$ for $t \in [1, \T_u-1]$ and $B_t = \frac{1-P_{tv}^{G}}{1-P_{tv}^{G'}}$ for $t \in [\T_u, \T_v-1]$. In either case, $B_t\leq 1$, since $P^G_{tu} \geq P^{G'}_{tu}$ and $P^G_{tv} \geq P^{G'}_{tv}$.

Product $C$ contains terms of nodes picked at $\T_v$, hence $C = \prod_{j: p_{\T_v j}=1}\frac{P_{\T_v j}^G}{P_{\T_v j}^{G'}} = \frac{P_{\T_v v}^G}{P_{\T_v v}^{G'}} = \exp(\epsilon')$ and $D=1$ because other than $u$ (already removed) and $v$ (in $C$), all other node's probabilities must be the same in $G$ and $G'$.


Then $P = \prod_{t = 1}^{\T_v-1}(A_t\times B_t)\times C \times D \leq \exp(2\epsilon')$. \\

\textbf{Case 2.b} $G' + (u, v) = G$

Using similar argument as in \textbf{Case 1.b}, we can calculate the values of $A_t, C, D$, with $\exp(\epsilon')$ is reversed to $\exp(-\epsilon')$, since we reverse the roles of $G$ and $G'$.
Then, $A_t = 1$ for all $t\neq \T_u$, and $A_{\T_u}=\frac{P^{G}_{\T u}}{P^{G'}_{\T u}}=\exp(-\epsilon') \leq 1$.

Similarly, $C = \prod_{j: p_{\T_v j}=1}\frac{P_{\T_v j}^G}{P_{\T_v j}^{G'}} = \frac{P_{\T_v v}^G}{P_{\T_v v}^{G'}} = \exp(-\epsilon') \leq 1$ and $D=1$.

Now we analyze $\prod_{t=1}^{\T_v-1}B_t$:

\begin{eqnarray*}
  \prod_{t=1}^{\T_v-1}B_t &=& \prod_{t=1}^{\T_u-1}\frac{(1-P_{tu}^G)(1-P_{tv}^{G})}{(1-P_{tu}^{G'})(1-P_{tv}^{G'})}  \times \prod_{t=\T_u}^{\T_v-1}\frac{1-P_{tv}^{G}}{1-P_{tv}^{G'}} \\
                          &=& \prod_{t=1}^{\T_u-1}(E_{tu} \times E_{tv}) \prod_{t=\T_u}^{\T_v-1}(E_{tv})\\
                          &\leq& \prod_{t=1}^{\T_u-1}\exp\left(\frac{4\epsilon'}{1-e^{-1}}p_t(G)\right) \times \prod_{t'=1}^{\T_v-\T_u-1}\exp\left(\frac{2\epsilon'}{1-e^{-1}}q^v_{t'}(G)\right) \\
                          &=& \exp\left(\frac{4\epsilon'}{1-e^{-1}}\sum_{t=1}^{\T_u-1}p_t(G)\right) \times \exp\left(\frac{2\epsilon'}{1-e^{-1}}\sum_{t'=1}^{\T_v-\T_u-1}q^v_{t'}(G)\right) \\
                          &\leq& \exp\left(\frac{4\epsilon'\ln{(1/\delta')}}{1-e^{-1}}\right)\times \exp\left(\frac{2\epsilon'\ln{(1/\delta')}}{1-e^{-1}}\right)\\
                          &=& \exp\left(\frac{6\epsilon'\ln{(1/\delta')}}{1-e^{-1}}\right)\\
                          &=& \exp\left(\frac{6\epsilon'\ln{(4/\delta)}}{1-e^{-1}}\right),
\end{eqnarray*}
where the first inequality is the result of Lemma~\ref{lemma:Etj} and the last inequality is the result of Lemma~\ref{lemma:process-phases}.

Then $P = \prod_{t = 1}^{\T_v-1}(A_t\times B_t)\times C \times D \leq \exp\left(\frac{6\epsilon'\ln{(4/\delta)}}{1-e^{-1}}\right)$ with probability at least $1-\delta/2$. 

Using the same argument as in Algorithm~\ref{algo:parallel} and substituting $\epsilon = \frac{24\epsilon'\ln{(4/\delta)}}{1-e^{-1}}$, the sequence $\{\pi_0, \pi_1, \ldots\}$ is $(\epsilon/4, \delta/2)$-differentially private. For all $t$, we only care about non-empty $\pi_t$ where at least 1 node is removed at $t$. Retaining all non-empty $\pi_t$ preserves the $(\epsilon/4, \delta/2)$-differential privacy, as we treat the process of dropping all empty $\pi_t$ is a post-process of the private sequence $\{\pi_0, \pi_1, \ldots\}$ as Post-Processing Proposition. 

Next, we prove that the sequence $\{\widehat{\rho(S_0)}, \widehat{\rho(S_0)}, \ldots\}$ is $(\epsilon/4, \delta/2)$-differentially private. Since the modification of an edge of the input graph will change the value of each $\rho(S_i)$ by at most $1/|S_i|$, using Laplace mechanism, adding a noise with magnitude of $4\ln{n}/(|S_i|\epsilon)$ satisfies $\epsilon/(4\ln{n})$-differential privacy for each $\widehat{\rho(S_i)}$. 
Lemma~\ref{lemma:phase-iterations} proves that the sequence $\{\widehat{\rho(S_0)}, \widehat{\rho(S_0)}, \ldots\}$ has at most $\ln{n}$ elements with probability at least $1-n^2 \geq 1-\delta/2$ when $\delta \geq 2n^{-2}$.
The sequence $\{\widehat{\rho(S_0)}, \widehat{\rho(S_0)}, \ldots\}$ is $\epsilon/4$-differentially private with probability at least $1-\delta/2$ due to applying Composition Theorem on $\ln{n}$ elements. It implies the sequence is $(\epsilon/4, \delta/2)$-differentially private because any mechanism which is $\epsilon$-differentially private with probability at least $1-\delta$ is also $(\epsilon, \delta)$-differentially private~\cite{Vadhan2017}.

Finally, the sequence $\{S_0, S_1,\ldots\}$ is constructed by using the sequence $\{\pi_0, \pi_1,\ldots\}$ and the the information computed from $\{\widehat{\rho(S_0)}, \widehat{\rho(S_1)}, \ldots\}$. Because both sequences above are $(\epsilon/4, \delta/2)$-differentially private, applying the Composition Theorem, $\{S_0, S_1,\ldots\}$ is $(\epsilon/2, \delta)$-differentially private.
\end{proof}

\begin{theorem}
  \label{theorem:parallel-phases-privacy-full}
Algorithm~\ref{algo:parallel-phases} preserves $(\epsilon, \delta)$-differential privacy, with $\epsilon\in(0,1], \delta\in(2n^{-2},1)$.
\end{theorem}

\begin{proof}
  Lemma~\ref{lemma:phase-privacy} proves that the sequence $\{S_1, S_2, ...\}$ output by the \textbf{while} loop is $(\epsilon/2, \delta)$-differentially private. 
  For each $S_i$, for any pair of neighbor graphs $G\sim G'$, the value of any $\rho(S_i)$ on $G$ and $G'$ differ by at most 1, hence $\Delta_{\rho} = 1$. Using the Exponential Mechanism, the last commands release an output with $\epsilon/2$-differential privacy. Applying the Composition Theorem on the sequence $\{S_1, S_2, \ldots\}$ which is $(\epsilon/2,\delta)$-differentially private and the output of the Exponential Mechanism which is $\epsilon/2$-differentially private, Algorithm~\ref{algo:parallel-phases} preserves $(\epsilon, \delta)$-differential privacy.

\end{proof}

\begin{theorem}
\label{theorem:parallel-phases-accuracy-full}
(Theorem~\ref{theorem:parallel-phases-accuracy})
Let $S^*$ be an optimal solution.
With probability at least $1-2/n$, the set $S$ output by Algorithm~\ref{algo:parallel-phases} satisfies $\rho(S)\geq \frac{\rho(S^*)}{4} -  \frac{160}{\epsilon}\ln(1/\delta)\ln(n)$ if $\delta \leq 1/e$.
\end{theorem}
\begin{proof}
Suppose $i$ is the first phase in which a node $v\in S^*$ is removed. Then, $S^*\subseteq S_i$. As in the proof of Theorem~\ref{theorem:approx-seq}, we have $\rho(S^*)\leq d_{S^*}(v)\leq d_{S_i}(v)$. From Lemma~\ref{lemma:parallel-node-maxdeg}, it follows, with probability at least $1-\frac{4\ln{n}}{n^2}$, that $d_{S_i}(v)\leq 4\widehat{\rho(S_i)} + \frac{4}{\epsilon'}\ln{n}\leq 4\rho(S_i) + \frac{128}{\epsilon}\ln{n} + \frac{4}{\epsilon'}\ln{n}$. Therefore, $\rho(S_i)\geq \frac{\rho(S^*)}{4} - \frac{32}{\epsilon}\ln{n} - \frac{1}{\epsilon'}\ln{n}$, with with probability at least $1-\frac{4\ln{n}}{n^2}\geq 1-1/n$.

As  in the proof of Theorem~\ref{theorem:approx-seq}, with probability at least $1-1/n$, we have $\rho(S)\geq \max_{i'}\rho(S_{i'}) - \frac{8}{\epsilon}\ln{n} \geq \rho(S_i) - \frac{8}{\epsilon}\ln{n}
\geq \frac{\rho(S^*)}{4} - \frac{40}{\epsilon}\ln{n} - \frac{24\ln{(4/\delta)}}{\epsilon(1-1/e)}{\ln{n}}
\geq \frac{\rho(S^*)}{4} - \frac{40}{\epsilon}\ln{n} - \frac{2.5*24\ln{(1/\delta)}}{1-1/e}\frac{1}{\epsilon}{\ln{n}}
\geq \frac{\rho(S^*)}{4} -  \left(\frac{40}{ln(1/\delta)}+\frac{60}{1-1/e}\right)\frac{1}{\epsilon}\ln(1/\delta)\ln{n}
\geq \frac{\rho(S^*)}{4} - \frac{160}{\epsilon}\ln(1/\delta)\ln{n}
$, 
and the Theorem follows.
\end{proof}
\subsection{Lower bound}
\begin{lemma}
  \label{lemma:lower-prob}
Let $G=(V, E)$ denote a graph with $|V|=n$ and $E=\emptyset$. 
Let $V$ be partitioned into sets $V_1,\ldots,V_N$, for $N=n/a$, with each $|V_i|=a$. For $i\in\{1,\ldots,N\}$, define $\mathcal{S}_i = \{S\subseteq V: S\cap V_i\neq\emptyset, |S|\leq a^2\}$.
Let $M(G)$ be an $\epsilon$-differentially private mechanism.
There exists an $i\in\{1,\ldots,N\}$ such that $\Pr[M(G)\in\mathcal{S}_i] \leq \frac{2a^3}{n}$
\end{lemma}

\begin{proof}
We now prove that there exists an $i\in\{1,\ldots,N\}$ such that $\Pr[M(G)\in\mathcal{S}_i] \leq \frac{2a^3}{n}$. Observe that 

\[
  \mathbb{E}\Big[|M(G)| \Big| |M(G)|\leq a^2\Big] 
  = \sum_{i=1}^N \mathbb{E}\Big[|M(G)\cap V_i| \Big| |M(G)|\leq a^2\Big]\leq a^2,
\]
which implies that there exists $i$ such that $\mathbb{E}\Big[|M(G)\cap V_i| \Big| |M(G)|\leq a^2\Big] \leq \frac{a^2}{N} = \frac{a^3}{n}$.


By the Markov's inequality, we have
$\Pr\Big[|M(G)\cap V_i| > \frac{n}{2a^3}\frac{a^3}{n} \Big| |M(G)|\leq a^2\Big] \leq \frac{2a^3}{n}$, 
which implies $\Pr[M(G)\cap V_i\neq\emptyset \Big| |M(G)\leq a^2] \leq \Pr[|M(G)\cap V_i|\geq 1\Big| |M(G)|\leq a^2] \leq \frac{2a^3}{n}$. Finally,

{\centering
{\scalebox{0.88}
{
$\begin{aligned}
  &\Pr[M(G)\in \mathcal{S}_i] \\
  = &\Pr[M(G)\cap V_i\neq\emptyset, |M(G)|\leq a^2]\\
  = &\Pr[M(G)\cap V_i\neq\emptyset\Big| |M(G)|\leq a^2] \Pr[|M(G)|\leq a^2] \\
  \leq &\frac{2a^3}{n}
\end{aligned}$}}

}
\end{proof}

\subsection{Map-Reduce Implementation}
\label{sec:mapr}
\begin{algorithm}
  \caption{\mralgo{}$(G, \epsilon, \delta)$\\
    \textbf{Input:} $G\in\Graphs$, privacy parameters $\epsilon, \delta$.\\
    \textbf{Output:} A subset $S\subseteq V(G)$ 
  }
  \label{algo:mapr-parallel-phases}
  \begin{algorithmic}[1]
    \STATE For all edges $(u, v)\in E(G)$, emit $\langle u, v \rangle$ and initialize variables as in \phasealgo{} 
    
    \WHILE {$|S_i| \neq \emptyset $}

    \STATE /*Start a new phase $i$*/

    \STATE \textbf{Reduce sub-phase}: Input $\langle u; v_1, v_2,\ldots \rangle$

    \IF {$u \in \{v_1, v_2, \ldots\}$}

    \STATE Ignore node $u$, it was sampled out in the last phase

    \ELSE

    \STATE Emit $\langle u; \{v_1, v_2, \ldots\} \rangle$ 
    \STATE Emit $\langle S_i; u\rangle$
    \STATE Emit $\langle \rho(S_i); deg_{S_i}(u)= |\{v_1, v_2, \ldots\}|\rangle$

    \ENDIF

    \STATE \textbf{Reduce sub-phase}: Input $\langle S_i; u_1, u_2,\ldots \rangle$

    \STATE $S_i = \{u_1, u_2, \ldots\}$

    \STATE \textbf{Reduce sub-phase}: Input $\langle \rho(S_i); deg_{S_i}(u_1), deg_{S_i}(u_2),\ldots \rangle$
    \STATE $\rho(S_i) = \sum_i deg_{S_i}(u_i) / 2$
    \STATE Calculate $\widehat{\rho(S_i)} \text{ and } T_i$ as in \phasealgo{}, line 12-13

    \STATE \textbf{Map phase sub-phase}: Input $\langle u, \{v_1, v_2, \ldots\} \rangle$ 
    \STATE Calculate $deg_{S_i}(u)= |\{v_1, v_2, \ldots\}|$

    \STATE Sample $T_u$ as in \phasealgo{}, line 11

    \IF {$T_u \leq T_i$}

    \STATE Emit $\langle u, u \rangle$ 

    \ELSE

    \STATE For $v \in \{v_1, v_2, \ldots\}$, emit $\langle v, u \rangle$

    \ENDIF
    
    \STATE $i = i + 1$

    \ENDWHILE
    
    \STATE Return $S_i$ from distinct $\{S_i\}$ with probability proportional to $e^{\epsilon\rho(S_i)/2}$ 
  \end{algorithmic}
\end{algorithm}

\begin{lemma}
  \label{lemma:map-reduce}
\mralgo{} preserves $(\epsilon, \delta)$-differential privacy, for $\epsilon\in(0,1], \delta\in(2n^{-2},1)$ and runs in $O(\ln n)$ phases, with probability at least $1-1/n^2$.
\end{lemma}
\begin{proof}
 We prove that for all $i$, the $i^{th}$ phase of \mralgo{} is equivalent to the $i^{th}$ phase of \phasealgo{}. The lemma then follows.

  In the first reduce sub-phase, with its input is every edge $(u, v)$, the sub-phase detects a removed node by looking for the node in its neighbor sets. It is because in lines 19-20, the algorithm marks a removed node in the previous phase by adding an edge that points to the node itself. \mralgo{} constructs three entities for each remained node $u$: the neighbor nodes of $u: \{v_1, v_2, \ldots\}$, assigning $u$ to candidate subgraph $S_i$ and the degree of $u$. 

  In the second and third reduce sub-phases, they gather all remained node $u$ and their degrees to construct the candidate subgraph $S_i$ and $\rho(S_i)$ as the results of the phase. They are the same as $S_i$ and $\rho(S_i)$ produced by Algorithm~\ref{algo:parallel-phases}.

  The map sub-phase takes input as pairs of node $u$ and $u$'s neighbors $\{v_1, v_2, \ldots\}$. The sub-phase calculates the degree of node $u$ with respect to the current nodes in phase $i$ and samples the removal time $T_u$ of each node $u$ as in line 11 of Algorithm~\ref{algo:parallel-phases}. It compares $T_u$ to $T_i$ (of phase $i$) to determine if $u$ is removed in this phase. If it is removed, the sub-phase only emits an self-edge to mark that the node is removed. It guarantees that in the next phases, there is no edge $(v, u)$ for any $v\neq u$ and hence, $u$ will not exist in the neighbor set of any node $v$. Otherwise, it emits all edge $(v, u)$ which will be used to construct the neighbor set of each node $v$ in the next phase.
\end{proof}

\clearpage
\section{Experiments}
\subsection{Experimental Setup}
\label{sec:setup}
In private algorithms, there are two parameters to control the privacy guarantee: $\epsilon$ and $\delta$. We can interpret $\epsilon$ as the upper bound on the absolute value of \textit{privacy loss}. While there is no strict range for the values of $\epsilon$, we are usually interested in small $\epsilon$, since it provides better privacy guarantee~\cite{dwork2011differential}. In the experiments, we test our algorithms with $\epsilon$ varying in the range $[ 0.5, 16]$, which is similar to the range of $\epsilon$ in~\cite{10.1145/2976749.2978318}. We infer $\delta$ as the ``upper bound on the probability of catastrophic failure (e.g., the entire dataset being published in the clear)''~\cite{Vadhan2017}. Hence, all applications must choose $\delta$ to be cryptographically negligible ($\delta \leq n^{-\omega(1)}$)~\cite{Vadhan2017}~\cite{dwork2011differential}. Since the networks we study have several hundreds of thousands of edges, we select $\delta \in \{10^{-6}, 10^{-9}\}$ for our experiments.

We perform each experiment (as a combination of algorithm with parameters) repeatedly 10 times to report the average of estimations. We must note that running multiple instances of the same algorithm is only for the sole purpose of performance analysis. In practice, multiple executions of the same private mechanism will degrade the privacy guarantee, in the worst case linearly to the number of executions~\cite{dwork:fttcs14}. 

\subsection{Experimental Results}
\label{sec:additional-results}
\begin{figure}[h]
  \includegraphics[width=0.5\columnwidth,keepaspectratio]{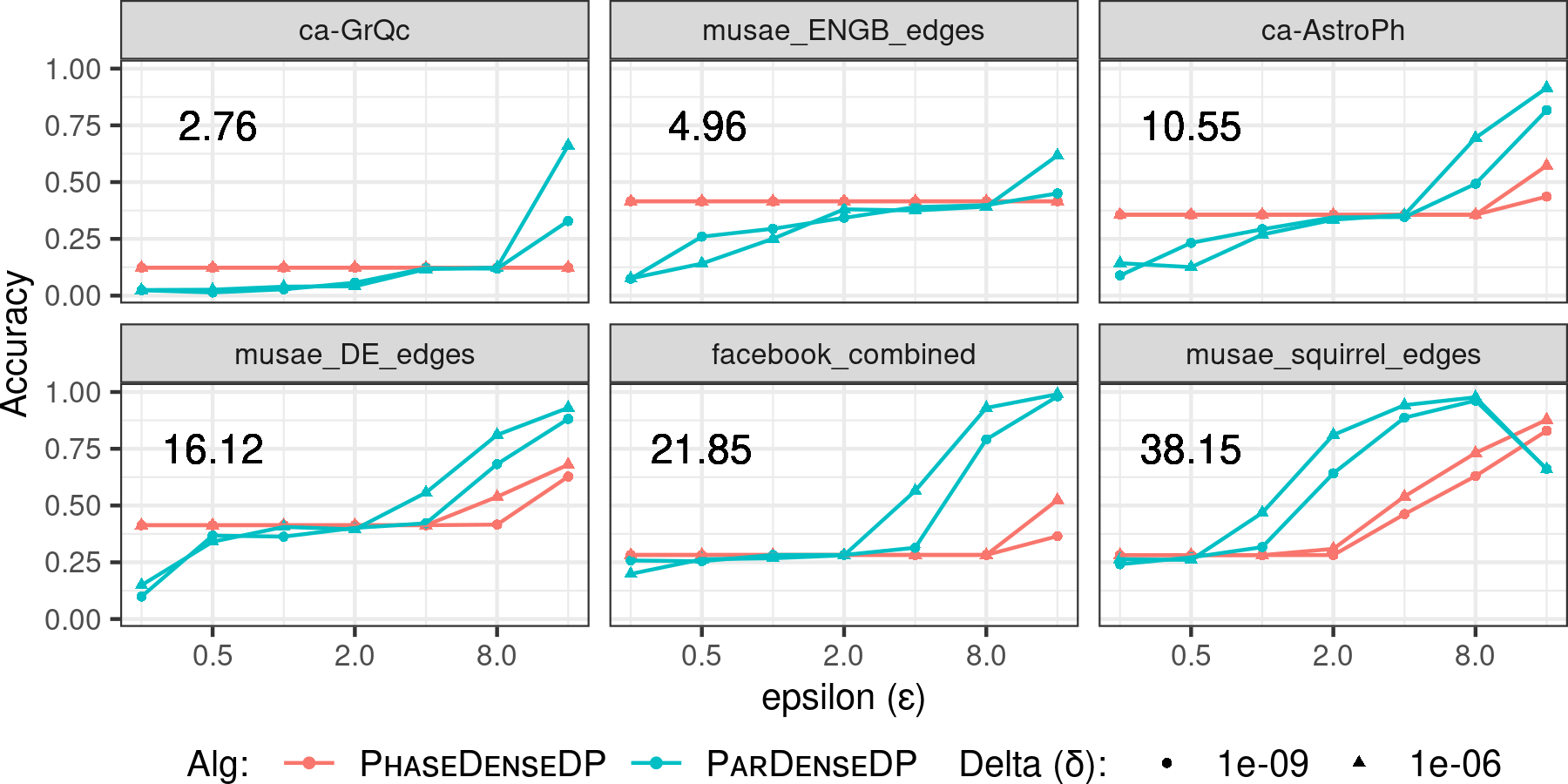}
  \centering
\caption{\textbf{Comparison of relative density between \palalgo{} and \phasealgo{}.}}
  \label{fig:bounded}
\end{figure}

\textbf{Performance of parallel algorithms.} 
Figure~\ref{fig:bounded} shows the accuracy comparison of \palalgo~and \phasealgo. In general, \palalgo~outperforms \phasealgo~at $\epsilon \geq 2$. For small $\epsilon$, neither algorithm has good accuracy, though \phasealgo~is slightly better. As we can see from Figure~\ref{fig:parIters}, \palalgo~is efficient in many real networks and settings, it is reasonable to choose it over \phasealgo~in practice.

%

\begin{table*}[]
\footnotesize
  \begin{tabular}{llllllll}
    \textbf{Network}        & \textbf{Nodes}  & \textbf{Edges}  & \textbf{Description}                          & \textbf{Network}          & \textbf{Nodes} & \textbf{Edges}  & \textbf{Description}                            \\
    ca-GrQc        & 5242   & 14496  & Collab. net. Arxiv Gen. Relativity   & deezer\_europe   & 28281 & 92752  & Social net. Deezer users from Europe   \\
    ca-HepTh       & 9877   & 25998  & Collab. net. Arxiv HE Phy. Theory    & musae\_DE        & 9498  & 153138 & Social net. of Twitch users (DE)       \\
    ca-HepPh       & 12008  & 118521 & Collab. net. Arxiv HE Phy.           & musae\_ENGB      & 7126  & 35324  & Social net. of Twitch users (ENGB)     \\
    ca-AstroPh     & 18772  & 198110 & Collab. net. Arxiv Astro Phy.        & musae\_facebook  & 22470 & 171002 & Facebook page-page net.                \\
    ca-CondMat     & 23133  & 93497  & Collab. net. Arxiv Cond. Matter      & musae\_FR        & 6549  & 112666 & Social net. of Twitch users (FR)       \\
    email-Enron    & 36692  & 183831 & Email net. from Enron                & musae\_git       & 37700 & 289003 & Social net. of Github developers.      \\
    loc-gowalla    & 196591 & 950327 & Gowalla location based social net    & musae\_squirrel  & 5201  & 198493 & Wiki page-page net. on topic squirrel  \\
    loc-brightkite & 58228  & 214078 & Brightkite location based social net & musae\_crocodile & 11631 & 170918 & Wiki page-page net. on topic crocodile \\
    facebook       & 4039   & 88234  & Social circles from Facebook         & musae\_chameleon & 2277  & 31421  & Wiki page-page net. on topic chameleon \\
    lastfm\_asia   & 7624   & 27806  & Social net. LastFM users from Asia   & musae\_PTBR      & 1912  & 31299  & Social net. of Twitch users (PTBR)
  \end{tabular}
  \caption{Summary of networks in our experiments. All networks are undirected, with sizes range from thousands to hundreds thousands.~\cite{snapnets}}
  \label{tab:full-nets}
\end{table*}

\begin{figure*}[h]
  \includegraphics[width=\textwidth,keepaspectratio]{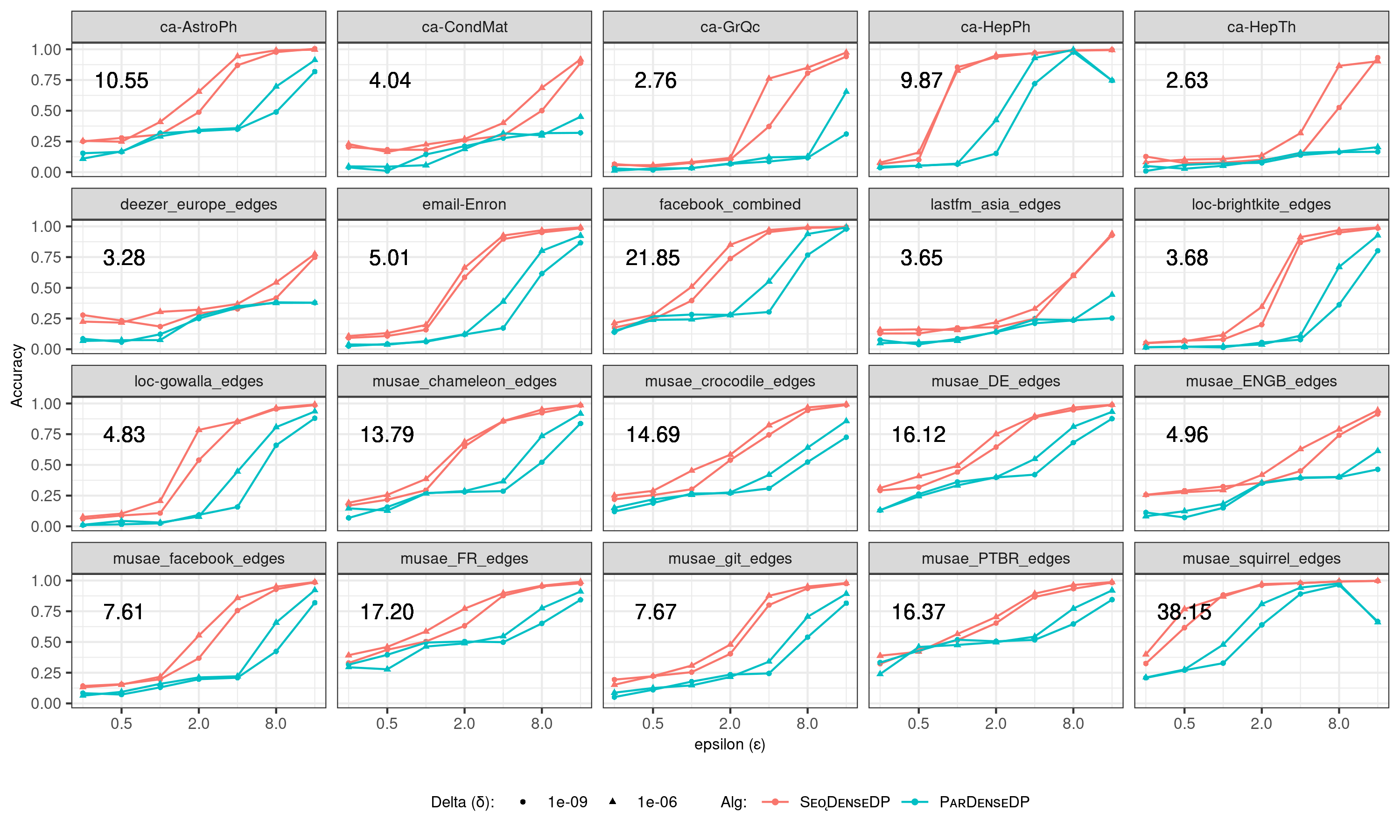}
  \centering
  \caption{\textbf{Accuracy of our private algorithms.} The accuracy is measured in terms of the relative density.  The number in each graph indicates the density of the graph as whole (which equals the average degree). An accuracy close to 1.0 means the private subgraph's density is as good as the non-private baseline's.}
  \label{fig:full-accuracy}
\end{figure*}

\end{document}